\documentclass[lettersize,journal]{IEEEtran}
\usepackage{amsmath,amsfonts}
\usepackage{algorithmic}
\usepackage{algorithm}
\usepackage{array}
\usepackage{textcomp}
\usepackage{stfloats}
\usepackage{url}
\usepackage{verbatim}
\usepackage{graphicx}
\usepackage{cite}
\usepackage{bbm}
\usepackage{subfigure}
\usepackage{color}

\begin{document}

\title{Age of Information in Gossip Networks: \\ A Friendly Introduction and Literature Survey}

\author{Priyanka Kaswan \quad Purbesh Mitra \quad Arunabh Srivastava \quad Sennur Ulukus\thanks{The authors are with the Department of Electrical and Computer Engineering, at the University of Maryland, College Park, MD, 20742, USA. Emails: \{pkaswan, pmitra, arunabh, ulukus\}@umd.edu.}}

\maketitle

\begin{abstract}\label{abstract}
Gossiping is a communication mechanism, used for \emph{fast} information dissemination in a network, where each node of the network randomly shares its information with the neighboring nodes. To characterize the notion of \emph{fastness} in the context of gossip networks, age of information (AoI) is used as a timeliness metric. In this article, we summarize the recent works related to \emph{timely gossiping} in a network. We start with the introduction of randomized gossip algorithms as an epidemic algorithm for database maintenance, and how the gossiping literature was later developed in the context of rumor spreading, message passing and distributed mean estimation. Then, we motivate the need for timely gossiping in applications such as source tracking and decentralized learning. We evaluate timeliness scaling of gossiping in various network topologies, such as, fully connected, ring, grid, generalized ring, hierarchical, and sparse asymmetric networks. We discuss age-aware gossiping and the higher order moments of the age process. We also consider different variations of gossiping in networks, such as, file slicing and network coding, reliable and unreliable sources, information mutation, different adversarial actions in gossiping, and energy harvesting sensors. Finally, we conclude this article with a few open problems and future directions in timely gossiping.
\end{abstract}


\section{Introduction}\label{sec: intro}
\begin{figure}[t]
\subfigure[Connected networks of vehicles.]{
\centerline{\includegraphics[scale=0.4]{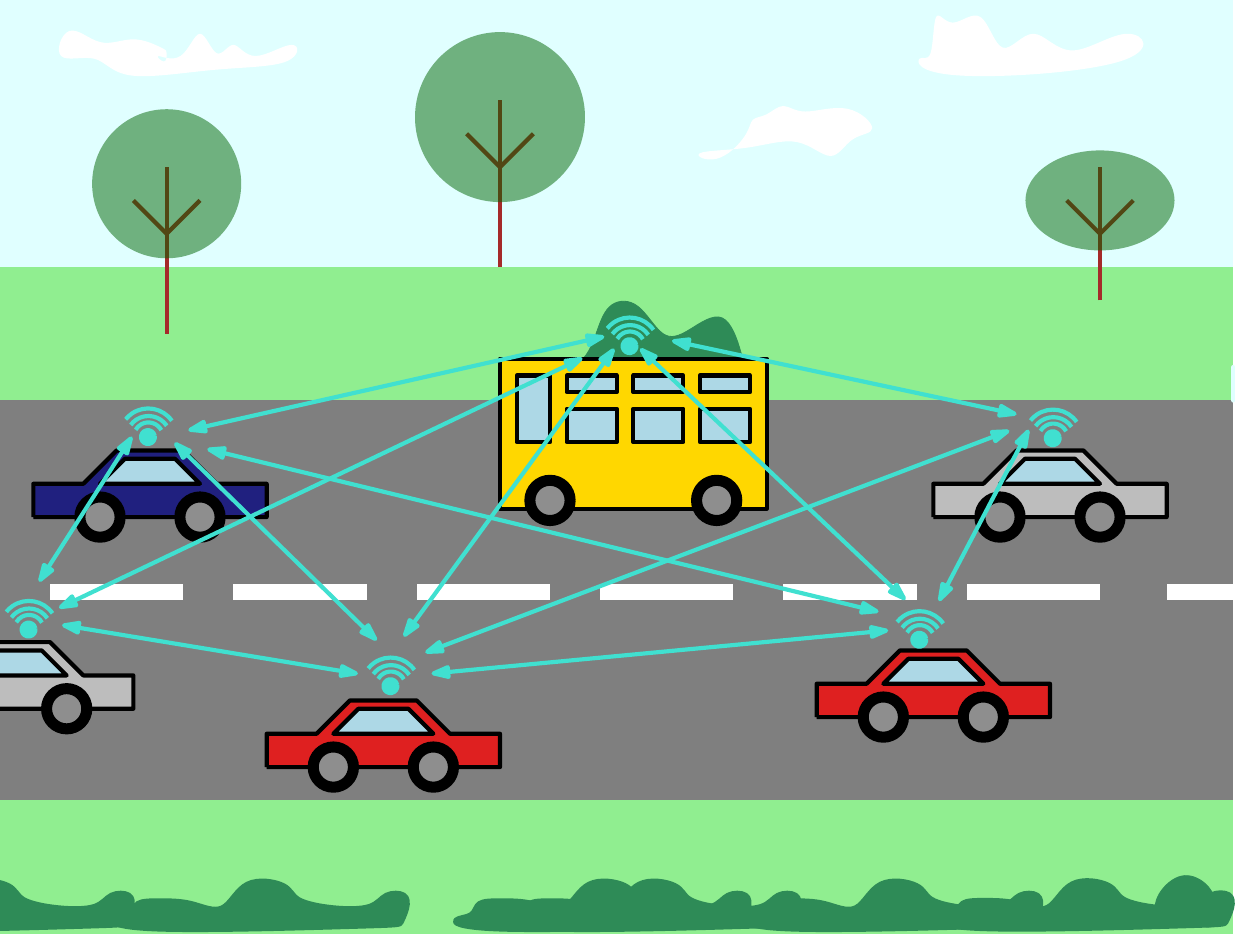}}
\label{fig: cars_gossip}}
\subfigure[Connected network of machines/devices.]{
\centerline{\includegraphics[scale=0.75]{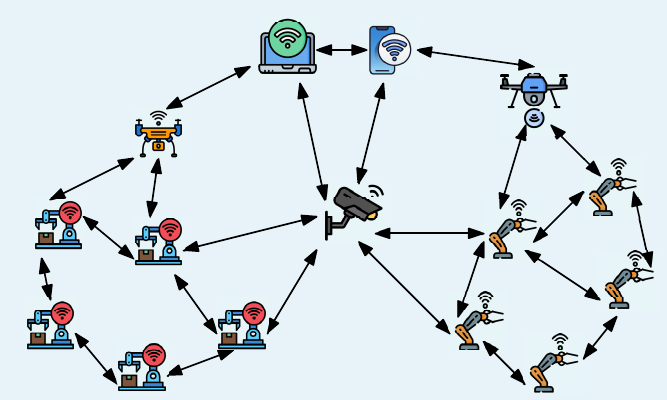}}
\label{fig: iot_gossip}}
\caption{Examples of real-world connected networks.}
\label{fig: real_gossip_net}
\end{figure}

In this review article, we discuss goal-oriented applications of gossip networks for time-sensitive information. An example of such applications is an autonomous driving system, where timely communication with nearby connected devices, such as other cars in the vicinity, sensors, infrastructure and even smartphones, is crucial to accurately perform driving actions and avoid accidents. A parallel example is a smart factory environment with human, robot, drone, camera collaborative system to safely and effectively perform manufacturing; see Fig.~\ref{fig: real_gossip_net}. Another application is remote surgery, where a doctor performs surgery on a patient using a remote surgical system, even though they are not physically present in the same location. The absence of real-time surgical data has to be taken into consideration to minimize the chance of inaccuracies in the surgical procedure. In such systems, there is a source, which has some time-varying information that is of interest to a user or a group of users. A user would like to track the time-varying information at the source as closely as possible in real-time to achieve its goal. However, there are network limitations, such as processing delays in the buffer queue, that prevent the user from  tracking the source arbitrarily closely, even with the high data rates afforded in the state-of-the-art communication systems.

\begin{figure}[t]
\centerline{\includegraphics[scale=0.8]{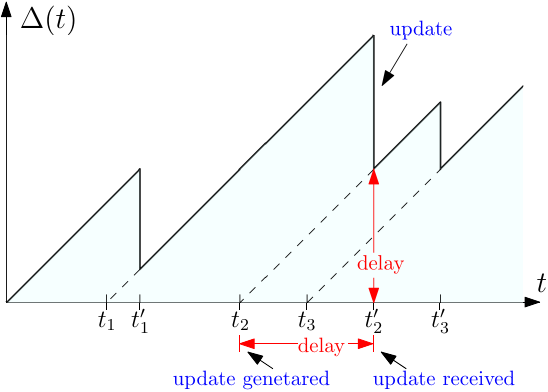}}
\caption{A sample path of the AoI process.}
\label{fig: AoI_sample}
\end{figure}

As a consequence of such limitations, the information at the user becomes stale if timely updates from the source are not received. Hence, it is crucial to optimize the communication parameters such that minimum staleness is maintained in the system. To achieve that, we need a metric that captures this staleness of the information at a user in a time-sensitive application. One such metric, proposed in the literature, is the age of information (AoI). For the latest information packet present at a user node at time $t$ that has the generation time of $u(t)$ at the source, the instantaneous AoI $\Delta(t)$ at the user is defined as $\Delta(t)=t-u(t)$. This simple metric essentially indicates how long ago a user's current packet was generated at the source. The AoI of a user increases at a unit rate as time progresses, until it receives a new packet from the source with a different generation time; see Fig.~\ref{fig: AoI_sample}. Ideally, a user would want $\Delta(t)=0$ for all $t$. However, this is not possible to achieve due to the limitations mentioned before. Therefore, it is desired to keep the AoI as low as possible. Since AoI is a time-dependent quantity, most literary works focus at optimizing either the time-average age, or peak-age, or some other statistical property of the age process at the user. Over an interval $[0,T]$ with large $T$, the average age is defined as
\begin{align}\label{eqn: avg_age}
    \Delta=\limsup_{T\to \infty}\frac{1}{T} \int_0^{T} \Delta(t) d t.
\end{align}
Graphically, the time-average age is the area under the saw-tooth curve in Fig.~\ref{fig: AoI_sample}, normalized by the interval of observation.

The idea of AoI was first introduced in \cite{kaul11AoI} in the context of vehicular networks and was generalized in the context of communication systems in \cite{kaul2012real}. Before going into the details of AoI formulation and its applications, it is crucial to motivate the use of AoI as a metric, and reason why we cannot rely on more traditional metrics such as delay and throughput that have been studied for decades in the literature \cite{kanodia2001distributed, gamal2004throughput, bisnik2006delay, li2010throughput, lu2013scaling}. To that end, \cite{kaul2012real} explains the novelty of AoI metric and its relevance to different time-sensitive applications. \cite{kaul2012real} explains this through an example, 
where in a vehicle, sensor measurements generated by various sensors are aggregated into a status update message and queued while they wait to be serviced by the car radio and transmitted to other cars. The radio interface is a simple first-come-first-served (FCFS) $M/M/1$ queue system with arrival rate of update packets as $\lambda$ and service rate $\mu$. \cite{kaul2012real} finds the server utilization $\rho=\frac{\lambda}{\mu}$ that minimizes the average age $\Delta$ for a fixed service rate $\mu$, by varying the arrival rate $\lambda$. It turns out that the optimal age is achieved with $\rho=0.53$, i.e., the server remains idle about $47\%$ of the time, such that the $\lambda$ biases the server towards being busy only slightly more than idle. Clearly to maximize the traditional metric of throughput, we would desire $\rho$ to be closer to $1$. However, since we are keeping the server always busy, we cause the queue to be backlogged with status update messages, leading to messages getting very stale or outdated by the time of their delivery. On the other hand, to minimize the traditional metric of delay, we would want $\rho$ to be close to $0$, since a low update rate leading to an empty queue lets a message get serviced right away. However, in this case, while keeping the queue empty, the other cars do not receive updates from this vehicle frequently enough, causing them to have outdated information about this vehicle, (voice communications are examples of such low delay low throughput applications). 

\begin{figure*}[t]
\subfigure[Low delay, low throughput.]{\centering
\includegraphics[scale=0.8]{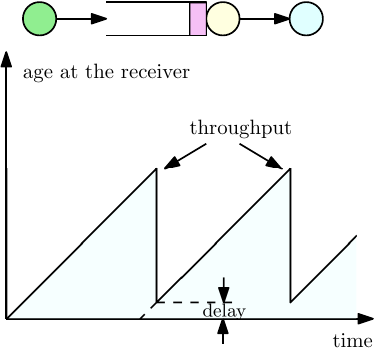}
\label{fig: ldlt}}
\hfill
\subfigure[High delay, high throughput.]{\centering
\includegraphics[scale=0.8]{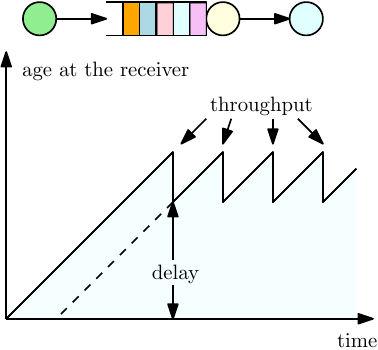}
\label{fig: hdht}}
\hfill
\subfigure[Medium delay, medium throughput.]{\centering
\includegraphics[scale=0.8]{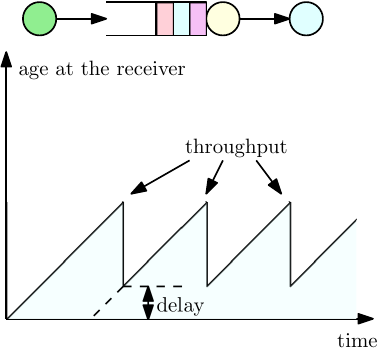}
\label{fig: mdmt}}
\caption{Delay versus throughput versus age trade-off in a queuing system.}
\label{fig: delay_tp_age}
\end{figure*}

This is explained pictorially in Fig.~\ref{fig: delay_tp_age}. In this figure, the sharp (vertical) decreases in the age denote the updates. Therefore, many such sharp decreases indicate a high update rate, hence a high throughput (as in Fig.~\ref{fig: hdht}). On the other hand, the height of the age right after an update has taken place denotes the delay that the update packet has experienced. Therefore, an age curve that decreases down to a small vertical height indicates a low delay (as in Fig.~\ref{fig: ldlt}). Finally, the normalized area under the age curve denotes the average age. Therefore, a smaller area for the same duration indicates a low average age (as in Fig.~\ref{fig: mdmt}). We observe that, generally, a low delay as in Fig.~\ref{fig: ldlt} comes at the cost of low throughput, and a high throughput as in Fig.~\ref{fig: hdht} comes at the cost of high delay, and a medium-throughput medium-delay point as in Fig.~\ref{fig: mdmt} yields a better age performance. Essentially, AoI is a complicated joint function of throughput and delay, and obtaining a good AoI outcome is about capturing a good trade-off point between jointly achievable throughput and delay. 

In recent years, various papers have explored age-optimal policies in a wide range of contexts, such as, queuing networks, energy harvesting systems, scheduling problems, UAV systems\cite{abbas2023comprehensive}, web crawling \cite{dhenakaran2011web}, remote estimation \cite{yatesJSACsurvey, kosta17AoIbook, sun2022age}, and so on. For the purposes of this article, we focus on the age of information in the context of a gossip network, which sometimes is called, the \emph{age of gossip}. 

Before we proceed to discuss the motivation of timeliness in gossiping in the next section, we take note of another metric: the \emph{version age of information} \cite{melih2020infocom, Abolhassani21version, yates21gossip}, useful for systems where update packets do not carry a generation timestamp. Instead, information packets generated at the source are marked by version numbers, that increment in steps of $1$ each time the source gets a new update. If $N_E(t)$ denotes the version number corresponding to the current state of the event and $N_i(t)$ denotes the version number of the information about the event present at node $i$, then the instantaneous version age of information at node $i$ is defined as $X_i(t)=N_E(t)-N_i(t)$, where $N_E(t)$ increments by one every time the event gets updated. Naturally, in this case, the user nodes wish to have access to the latest version present at the source, but unlike the traditional age of information which increases at unit rate, version age does not change at a user node in absence of a source update or a newer version update packet arrival at the user node. Additionally, the version age remains $0$ when the source and the node both have the same information, whereas AoI keeps increasing linearly with time if the source does not transmit any new information, thus penalizing the network even when there is no actual necessity of transmission. Evidently, the version age is particularly useful where information changes as per some counting process, such as, Poisson renewals, whereas, AoI is more suitable for continuous-time information dynamics, such as, temperature data.

In this review article, we mainly focus on AoI and its variant version age in the context of timely source tracking in a gossip network. First, we motivate the necessity of timeliness in a gossip network. We define the age of a gossip network and derive closed-form expressions for it. Then, we look into the growth of age as a function of the network size (age scaling) for different symmetrical gossip network structures, such as, fully connected, ring, grid, generalized ring, and hierarchical networks. We summarize the improvements based on different age-aware gossiping techniques. We discuss higher order moments of the age process. We also discuss the fairness optimization in timely operation of sparse asymmetric gossip networks. Then, we summarize works related to several variations of timely gossiping, such as, gossiping with file slicing and network coding, reliable and unreliable sources, information mutation, jamming and timestomping adversary, and gossiping with an energy harvesting sensor. Finally, we list several open problems and possible future directions.

\section{Timeliness in Gossip Networks}\label{sec: timely gossip}
\begin{figure}[t]
\subfigure[Fully connected network.]{
\centering{\includegraphics[scale=0.38]{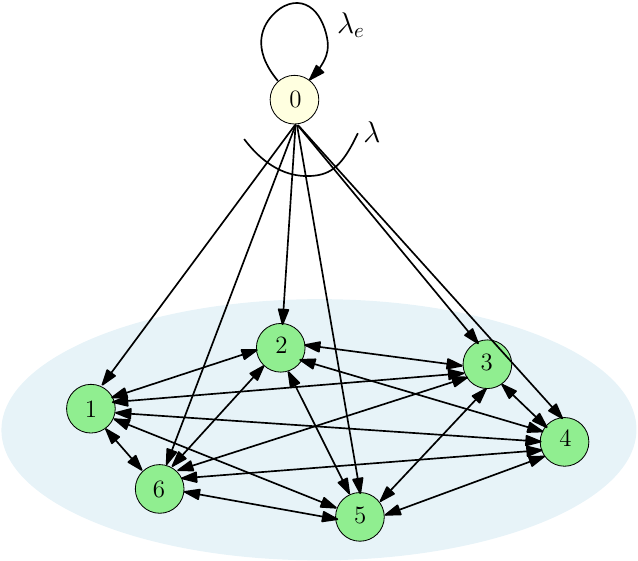}}
\label{fig: fc_network}
}\hfill
\subfigure[Ring network.]{
\centering{\includegraphics[scale=0.38]{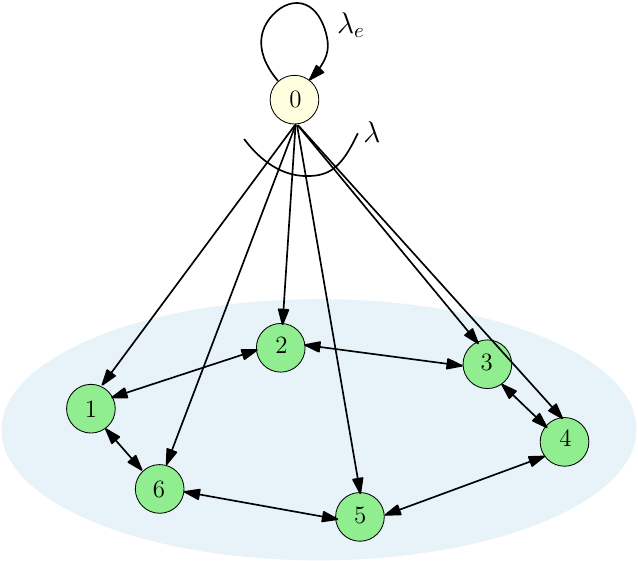}}
\label{fig: rc_network}}
\caption{Two example gossip networks consisting of $n=6$ nodes.}
\label{fig: simple_gossip_net}
\end{figure}

Gossiping is a fast and distributed information sharing mechanism in a network, where each node of the network randomly communicates to its neighboring nodes and spreads information; see Fig.~\ref{fig: simple_gossip_net}. Such gossip networks are commonly used for various resource-constrained, goal-oriented applications, where centralized scheduling does not offer simple scalable solutions. Examples of such applications include internet of things (IoT) networks \cite{chettri2019comprehensive, swamy2020empirical}, dense sensor networks \cite{akkaya2005survey, villalba2009routing, he2020distributed},
mobile ad-hoc networks \cite{tarique2009survey,ruiz2015survey}, content distribution networks \cite{li2007peer,el2011building, coileain2015accounting}, autonomous driving networks \cite{raza2019survey}, decentralized learning networks \cite{hegedus2019gossip, hegedus2021decentralized}. In high risk applications, such as, autonomous driving, it is crucial to have efficient communication with other cars in the vicinity and many sensors of the car to avoid road accidents. Such networks with large scale connectivity call for simple gossip based algorithms, where nodes arbitrarily exchange information with their neighboring nodes while being oblivious to the overall dynamics of the system, thereby causing information to spread like a
gossip or rumor. Indeed, hyper-connectivity among all humans and machines is one of the key promises of emerging sixth generation (6G) communication standards~\cite{lee23_6G}, where gossip networks will play a pivotal role. 

In the literature, the idea of gossip protocols was first introduced in \cite{demers87epidemic_gossip} as an epidemic algorithm for clearinghouse database maintenance. \cite{demers87epidemic_gossip} shows that randomized gossip algorithms are efficient in ensuring that every update in the system is eventually reflected in all the databases, thus, maintaining a consistency in the network. In the subsequent studies \cite{pittel_87_gossip, vocking2000, Minsky02cornellthesis}, the efficiency of gossip algorithms in rumor dissemination was analyzed. It is worth noting here that different works in the literature mention gossiping as peer-to-peer (P2P) or device-to-device (D2D) or machine-to-machine (M2M) communications, differing in semantics only. \cite{vocking2000} shows that a single rumor can be spread to $n$ nodes in $O(\log n)$ rounds. This order performance can be further extended to multiple rumors via network coding techniques~\cite{deb2006AlgebraicGossip, devavrat2006, Sanghavi2007GossipFileSplit}, which allow transmitting multiple pieces of information, encoded in a single message and decoding them intelligently to retrieve the desired information. \cite{deb2006AlgebraicGossip} studies dissemination time of $k$ messages in a large network of $n$ nodes with gossip protocols based on random linear coding (RLC), random message selection (RMS), and sequential dissemination. \cite{deb2006AlgebraicGossip} shows that RLC-based protocol has superior performance and has $ck + O(\sqrt{k}\log k \log n)$ dissemination time in complete graphs. \cite{devavrat2006} further extends the result to arbitrary graphs. In \cite{Sanghavi2007GossipFileSplit}, a file is split into $k$ pieces for faster dissemination in a large network of $n$ nodes. \cite{Sanghavi2007GossipFileSplit} shows that a dissemination time of $O(k+\log n)$ is achievable by a hybrid piece selection protocol, named the INTERLEAVE protocol. In \cite{boyd2005gossip,shah08monograph}, gossiping protocols for distributed mean estimation were analyzed. Such protocols are used for distributed multi-agent optimization and consensus problems~\cite{nedic2009distributed, nedic2010constrained,nedic2009distributed_avg}. 

The works mentioned so far consider the total dissemination time of a static message in the network as the performance metric. However, in real-time scenarios, data dynamics is time-dependent and asynchronous. In distributed databases such as Amazon DynamoDB \cite{amazondynamo} and Apache Cassandra~\cite{Cassandra}, the database nodes use gossip protocols to keep their information fresh. In Cassandra, cluster metadata at each node is stored in endpoint state, which tracks the version number or timestamp of the data. During a gossip exchange between two nodes, the version numbers of the data at the two nodes are compared, and the node with the older version discards its data, replacing it with the more up-to-date data of the other node. Thus, old information is often discarded by the nodes before it spreads to the whole network and only timely information is kept. 

Another application, which necessitates timely information dissemination in a network, is decentralized machine learning~\cite{hegedus2019gossip, hegedus2021decentralized}. Unlike centralized learning, decentralized learning is a method where the training data for learning a model is distributed across different devices. In this method, the devices themselves train local models with their available data and communicate their own model to their neighboring devices for averaging or mixing after exponential time intervals (Poisson arrivals). The model training and the model mixing can run asynchronously, which makes the setting equivalent to a dynamic information gossip model. Model mixing is essential to convergence of the overall network, especially with non-i.i.d.~distributed data. The memoryless property of the exponential distribution allows easier convergence analysis by discrete-time model formulation. The analysis in \cite{lian2018asynchronous,ram2009asynchronous,jin_scale_DDL,glasgow_adosd} shows that for guaranteed model convergence of a device, the version difference between the model available at the device and the global consensus version must be bounded by a constant and the rate of convergence becomes faster as the timeliness of the network improves.

These examples show that the total dissemination time is not an adequate metric to study in timely dissemination of dynamically changing information in a network. Rather, some metric that can incorporate information freshness for dynamic data, such as age or version age, is more suitable.

\section{Age of Gossip}\label{sec: age of gossip}
Most works prior to \cite{yates21gossip, Yates21gossip_traditional} studied the age for specific network topologies such as a simple transmitter-receiver pair. However, the saw-tooth curves become complex for arbitrary network topologies, making their analysis difficult. \cite{yates_realtime_multisrc} first demonstrated the application of stochastic hybrid systems (SHS) technique based on the model of \cite{hespanhashs} for the analysis of age processes. In the subsequent studies, the SHS characterization became crucial in analyzing age in gossip networks.

\subsection{Stochastic Hybrid System Modelling for Age Analysis}\label{sec: SHS_age}
An SHS is characterized by its state, which is partitioned into a discrete component $q(t) \in \mathcal{Q}=\{0,1, \ldots, m\}$ that evolves as a jump process and a continuous component $\pmb{x}(t)=\left[x_0(t) \cdots x_n(t)\right] \in \mathbb{R}^{n+1}$ that evolves according to a stochastic differential equation. Given the discrete set $\mathcal{Q}$ and a $k$ dimensional vector $\pmb{z}(t)$ of independent Brownian motion processes, the stochastic differential equation is expressed as
\begin{align}\label{eqn: stochatic_diff}
\dot{\pmb{x}}=f(q, \pmb{x}, t)+g(q, \pmb{x}, t) \dot{\pmb{z}},
\end{align}
where $f: \mathcal{Q} \times \mathbb{R}^{n+1} \times[0, \infty) \rightarrow \mathbb{R}^{n+1}$ and $g$ : $\mathcal{Q} \times \mathbb{R}^{n+1} \times[0, \infty) \rightarrow \mathbb{R}^{(n+1) \times k}$ are the mappings, and $\mathcal{L}=\left\{0, \ldots, \ell_0-1\right\}$ is a set of transitions, such that each $\ell \in \mathcal{L}$ is a discrete transition/reset map $\phi_{\ell}: \mathcal{Q} \times \mathbb{R}^{n+1} \times[0, \infty) \rightarrow \mathcal{Q} \times \mathbb{R}^{n+1}$. The corresponding transition
\begin{align}\label{eqn: transitions}
\left(q^{\prime}, \pmb{x}^{\prime}\right)=\phi_{\ell}(q, \pmb{x}, t)
\end{align}
occurs with transition intensity
\begin{align}\label{eqn: transition intensity}
\lambda^{(\ell)}(q, \pmb{x}, t), \quad \lambda^{(\ell)}: \mathcal{Q} \times \mathbb{R}^{n+1} \times[0, \infty) \rightarrow[0, \infty) .
\end{align}

In other words, the probability that the $\ell$th transition occurs in the interval $(t, t+d t]$ is $\lambda^{(\ell)}(q(t), \pmb{x}(t), t) d t$. When the system is in discrete state $(q, \pmb{x}(t))$, it evolves following \eqref{eqn: stochatic_diff}; but if it transitions from $q$ to $q^{\prime}$, the continuous state can have a discontinuous jump from $\pmb{x}$ to $\pmb{x}^{\prime}$, as given in \eqref{eqn: transitions}. Hence, the resulting process $\pmb{x}(t)$ has piecewise continuous sample paths. 

Due to the broad nature of the SHS model, describing the processes $q(t)$ and $\pmb{x}(t)$ can be intricate and challenging. The strategy proposed in \cite{yates21gossip} is to establish test functions $\psi(q, \pmb{x}, t)$, with the expected value denoted as $\mathbb{E}[\psi(q(t), \pmb{x}(t), t)]$, that can be assessed utilizing the method outlined in \cite[Thm.~1]{yates21gossip}. In the subsequent section, we begin by demonstrating this approach, specifically for the version age of information metric.

\subsection{SHS for Version Age of Information} \label{sec: SHS_version_age}
\begin{figure}[t]
\centerline{\includegraphics[scale=0.7]{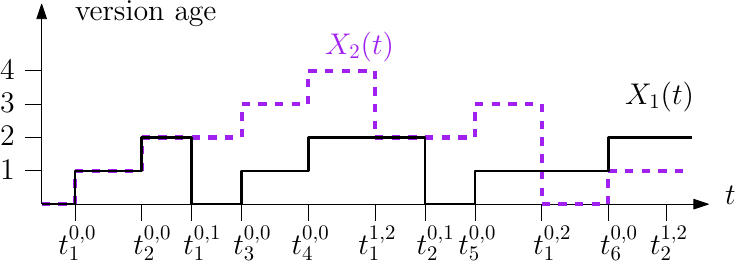}}
\caption{A sample path of version ages $X_1(t)$ and $X_2(t)$ in a gossip network of $2$ nodes, node $1$ and node $2$ (note that node $0$ is the source node). $t^{i,j}_k$ represents the $k$th update from node $i$ to node $j$.}
\label{fig: version_age_sample}
\end{figure}

We start with the exploitation of the SHS technique for the characterization of average version age in arbitrary networks, as explained in \cite{yates21gossip}. Consider an arbitrary network topology with $n$ user nodes $\mathcal{N}=\{1,2,\ldots,n\}$ and a source node $0$, where the source gets updated with newer versions according to $\lambda_{00}$ rate Poisson process such that all user nodes wish to have access to the latest possible version of this constantly updating information. The version age at node $i$ is denoted by $X_i(t)$, and node $i$ sends update packets to node $j$ as a $\lambda_{ij}$ rate Poisson process. Since the source always has the latest version of the information, $X_0(t)=0$ at all times. Each time the source gets updated, the age of node $i$ becomes $X_i'(t)=X_i(t)+1$. When node $i$ receives a packet from node $j$, the age of node $i$ becomes $X_i'(t)=\min\{X_i(t),X_j(t)\}$, since node $i$ keeps the fresher of the two packets and discards the staler one to improve its version age; see Fig.~\ref{fig: version_age_sample}. We wish to compute $\lim_{t \to \infty}E[X_i(t)]$, by employing the SHS method.

The continuous state of this SHS model is $\pmb{X}(t)=[X_1(t),\ldots,X_n(t)] \in \mathbb{R}^{n}$, which is a vector of the instantaneous ages at the $n$ nodes. The convenience of the SHS based version age characterization follows from the presence of a single discrete mode with trivial stochastic differential equation $\pmb{\dot X}(t)=\pmb{0}_n$, since the version age at nodes does not change between transitions. In the gossip network, the set of transitions $\mathcal{L}$ corresponds to the set of directed edges $(i,j)$, such that node $i$ sends updates to node $j$ on this edge according to a Poisson process of rate $\lambda_{ij}$, with $(0,0)$ denoting a source self-update, i.e.,
\begin{align}\label{eqn: updates}
    \mathcal{L}= \{(0,0)\} \cup \{(0,i):i \in \mathcal{N}\}\cup \{(i,j):i,j \in \mathcal{N}\},
\end{align}
where $(i,j)$ transition resets the state $\pmb{X}$  at time $t$ to $\phi_{i,j}(\pmb{X})=[X_1',\ldots,X_n']\in \mathbb{R}^{2n}$ post transition, such that
\begin{align}\label{eqn: reset map}
X_k'= \begin{cases}
    X_k+1, & i=0, j=0, k \in \mathcal{N} \\ 
    0, & i=0, k=j \in \mathcal{N} \\ 
    \min \left(X_i, X_j\right), & i \in \mathcal{N}, k=j \in \mathcal{N} \\
    X_k, & \text {otherwise}.
    \end{cases}
\end{align}

Frequently in the remaining part of this article, we will define a test function of the form $\psi:\mathcal{Q} \times \mathbb{R}^{n}\times [0,\infty) \to \mathbb{R}$ that is time-invariant, i.e., its partial derivative with respect to $t$ is $\frac{\partial\psi(q,\pmb{X},t)}{\partial t}=0$, such that we are interested in finding its long-term expected value $\lim_{t \to \infty} \mathbb{E}[\psi(q(t),\pmb{X}(t),t)]$. Since the test function only depends on the continuous state values $\pmb{X}$ due to time-invariance and single discrete mode, for simplicity, we will drop the inputs $q$ and $t$ and write $\psi(q,\pmb{X},t)$ as $\psi(\pmb{X})$. The test function $\psi(\mathbf{X}(t))$ has an extended generator $(L \psi)(\mathbf{X}(t))$ that satisfies Dynkin's formula,
\begin{align}\label{eqn: Dynkin}
\frac{d \mathrm{E}[\psi(\mathbf{X}(t))]}{d t}=\mathrm{E}[(L \psi)(\mathbf{X}(t))].     
\end{align}
Given the trivial stochastic differential equation $\pmb{\dot X}(t)=\pmb{0}_n$ and the time-invariance of the test function in our case, the extended generator of SHS for version age is given by \cite{yates21gossip},
\begin{align}\label{eqn: extended_gen}
    \left(L \psi_S\right)(\mathbf{X})=\sum_{(i, j) \in \mathcal{L}} \lambda_{i j}\left[\psi_S\left(\phi_{i, j}(\mathbf{X})\right)-\psi_S(\mathbf{X})\right]. 
\end{align}

To compute the expected version age at the nodes, \cite{yates21gossip} defines a test function $\psi_S(\mathbf{X})=X_S=\min_{j\in S}X_j$, for a set of nodes $S \subseteq \mathcal{N}$, which induces the process $\psi_S(\mathbf{X}(t))=X_S(t)=\min_{j\in S}X_j(t)$.
The effect on the test function of transition $(i, j)$ is
\begin{align}\label{eqn: test function}
    \psi_S\left(\phi_{i, j}(\mathbf{X})\right)=X_S^{\prime}=\min _{k \in S} X_k^{\prime},
\end{align}
which in accordance with \eqref{eqn: test function}, changes as follows,
\begin{align}\label{eqn: reset map 2}
X_S'= \begin{cases}
    X_S+1, & i=0, j=0 \\ 
    0, & i=0, j \in S \\ 
    X_{S\cup\{i\}}, & i \in N(S), j \in S \\ 
    X_S, & \text{otherwise},
    \end{cases}
\end{align}
where $N(S)$ is the set of updating neighbors of $S$,
\begin{align}\label{eqn: neighbors}
    N(S)= \bigg\{i\in \mathcal{N}\backslash S: \lambda_i(S)=\sum_{j \in S}\lambda_{ij}>0 \bigg\}.
\end{align}
The extended generator, therefore, becomes,
\begin{align}\label{eqn: extended_gen 2}
    \left(L \psi_S\right)(\mathbf{X})=&\lambda_{00}\left(X_S+1-X_S\right)+\sum_{j \in S} \lambda_{0 j}\left[0-X_S\right] \nonumber \\
    &+\sum_{i \in N(S)} \sum_{j \in S} \lambda_{i j}\left[X_{S \cup\{i\}}-X_S\right].
\end{align}

With the definitions $\mathbb{E}[X_S(t)]=v_S(t)$ and $v_S=\lim_{t \to \infty} v_S(t)$, substituting \eqref{eqn: extended_gen 2} into \eqref{eqn: Dynkin} gives, 
\begin{align}\label{eqn: derivative_eqn}
    \dot{v}_S(t)=&\lambda_{00} - v_S(t)\lambda_0(S)  \nonumber\\
    &+ \sum_{i\in N(S)} \lambda_i(S)[v_{S \cup\{i\}}(t)-v_S(t)]. 
\end{align}
As $t$ tends to $\infty$, setting $\dot{v}_S(t)=0$ yields,
\begin{align}\label{eqn: recursive equations}
v_S=\frac{\lambda_{00}+\sum_{i \in N(S)} \lambda_i(S) v_{S \cup\{i\}}}{\lambda_0(S)+\sum_{i \in N(S)} \lambda_i(S)}.
\end{align}

Equation \eqref{eqn: recursive equations} carries significant importance in the subsequent studies in timely gossip networks. \eqref{eqn: recursive equations} expresses the version age of a set of size $|S|$ in terms of the ages of sets of size $|S|+1$ by incorporating the neighboring nodes of set $S$ one by one. \eqref{eqn: recursive equations} also allows us to compute expected version age in very large networks of certain types of topologies, as we will see in the remaining sections of this article. Specifically, in symmetric fully connected network and symmetric ring network, by exploiting the symmetry, the equation will reduce to a summation in the case of fully-connected networks and a Riemann integral in the case of ring networks.

Before moving on to the next section, we would like to remark that in the case of the age of information instead of the version age of information, it has been shown in \cite{Yates21gossip_traditional} that for time-invariant test functions, the extended generator counterpart of \eqref{eqn: extended_gen} is
\begin{align}\label{eqn: extended_gen 3}
    \left(L \psi_S\right)(\mathbf{X})=1+ \sum_{(i, j) \in \mathcal{L}} \lambda_{i j}\left[\psi_S\left(\phi_{i, j}(\mathbf{X})\right)-\psi_S(\mathbf{X})\right]. 
\end{align}

Using similar steps as before, \eqref{eqn: extended_gen 3} results in the formula for the average age of information for a set $S$ as,
\begin{align}\label{eqn: recursion_AoI}
    v_S=\frac{1+\sum_{i \in N(S)} \lambda_i(S) v_{S \cup\{i\}}}{\lambda_0(S)+\sum_{i \in N(S)} \lambda_i(S)}.
\end{align}
Note that the formula in \eqref{eqn: recursion_AoI} can be obtained from \eqref{eqn: recursive equations} if we set $\lambda_{00}=1$. Thus, for the rest of the article, the results obtained for the version age of information metric can be readily applied to the traditional age of information metric by just setting the source self-update rate as $\lambda_{00}=1$.

\section{Age Scaling for Simple Network Topologies}\label{sec: age_scaling_simple}
\begin{figure}
    \centering
    \includegraphics[scale = 0.5]{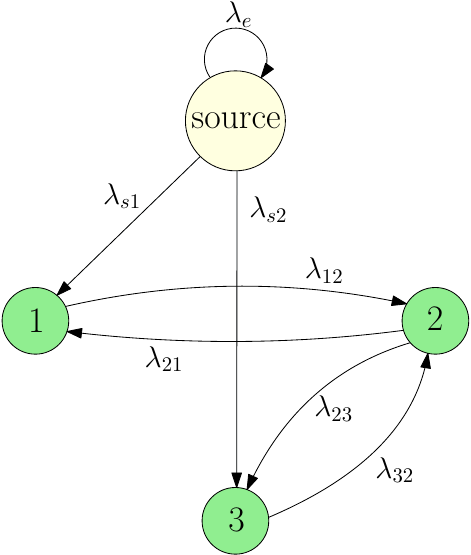}
    \caption{A toy example of a source node and three gossiping nodes.}
    \label{fig: toy example}
\end{figure}

In order to find the version age scaling for simple networks, we can use the recursive equations. However, the number of such equations that we need to solve recursively is exponential. In the following toy example, we demonstrate how to compute the expected ages at all nodes for a small network. Consider the network in Fig.~\ref{fig: toy example}. Here, the source sends updates to nodes $1$ and $3$; there is no direct communication link between nodes $1$ and $3$; and nodes $1$, $2$ and $2$, $3$ gossip. From \eqref{eqn: recursive equations}, we have 
\begin{align}\label{eqn: toy_example_age 0}
v_{\{1,2,3\}} = \frac{\lambda_e}{\lambda_{s1}+\lambda_{s2}}. 
\end{align}
Using this, we can find the version age of two-sized sets as,
\begin{align}
    v_{\{1,2\}} = \frac{\lambda_e + \lambda_{32}v_{\{1,2,3\}}}{\lambda_{s1} + \lambda_{32}},\label{eqn: toy_example_age 1-1}\\
    v_{\{2,3\}} = \frac{\lambda_e + \lambda_{12}v_{\{1,2,3\}}}{\lambda_{s2} + \lambda_{12}}.\label{eqn: toy_example_age 1-2}
\end{align}
Then, we can find the version age of single nodes as,
\begin{align}
    v_1 &= \frac{\lambda_e + \lambda_{21}v_{\{1,2\}}}{\lambda_{s1} + \lambda_{21}}, \label{eqn: toy_example_age 2-1}\\
    v_2 &= \frac{\lambda_e + \lambda_{12}v_{\{1,2\}} + \lambda_{32}v_{\{2,3\}}}{\lambda_{12}+\lambda_{32}},\label{eqn: toy_example_age 2-2}\\
    v_3 &= \frac{\lambda_e + \lambda_{23}v_{\{2,3\}}}{\lambda_{s3} + \lambda_{23}}.\label{eqn: toy_example_age 2-3}
\end{align}
That is, we calculate the version age of each node in the network using the recursive equation and by forming sets that are one-larger. For example, to calculate the age of node $2$, in \eqref{eqn: toy_example_age 2-2}, we write the age of set $\{2\}$ by the ages of one-larger sets $\{1,2\}$ and $\{2,3\}$. Then, to find the ages of these two-sized sets, in \eqref{eqn: toy_example_age 1-1} and \eqref{eqn: toy_example_age 1-2}, we write their ages by the age of one-larger set $\{1, 2, 3\}$. Since there is no one-larger set to $\{1, 2, 3\}$, the recursive equation \eqref{eqn: recursive equations} directly gives its age as in \eqref{eqn: toy_example_age 0}. Now, \eqref{eqn: toy_example_age 0} will be inserted into \eqref{eqn: toy_example_age 1-1} and \eqref{eqn: toy_example_age 1-2}, which will be inserted into \eqref{eqn: toy_example_age 2-2} to find the age of node $2$. Note also, interestingly, that we did not evaluate $v_{\{1,3\}}$ here, since it does not appear in the calculation of the version age of any node.

Note that, in general, we may have as many equations as the number of subsets of $n$ nodes, which is exponential in $n$. In order to simplify the calculations for larger networks, we need to exploit the geometry of the network to reduce the number of recursive equations and simplify the calculations. The number of equations can be reduced to $O(n)$ in the special cases of the bidirectional ring network and the fully connected network.

\subsection{Age Scaling for a Fully Connected Network} \label{sec: age_scaling_simple-FC}
The fully connected network, as shown in Fig.~\ref{fig: fc_network}, has the highest density of connections, with each node sharing an edge with every other node in the network, as analyzed in \cite{yates21gossip}. Since the total gossip rate of each node is $\lambda$, each node gossips with every other node with rate $\frac{\lambda}{n-1}$. In order to find the version age of a single node in the network, we first notice that the version age of two sets of the same size is the same, and hence we can write the version age of set $v_S = v_{|S|}$. Hence, we obtain a sequence of version ages $v_1,v_2, \ldots, v_n$, where $v_n = \frac{\lambda_e}{\lambda}$. Then, the recursion in \eqref{eqn: recursive equations}, expressed only in terms of the subset size $j$ is given by (see also \cite[Eqn.~(10)]{yates21gossip}),
\begin{align} \label{yates-fc}
    v_j = \frac{\lambda_e + \frac{j(n-j)\lambda}{n-1} v_{j+1}}{\frac{j\lambda}{n} + \frac{j(n-j)\lambda}{n-1}}.
\end{align}
In principle, one can start from $v_n = \frac{\lambda_e}{\lambda}$, and work backwards $j=n-1, n-2, \ldots, 2, 1$ utilizing \eqref{yates-fc} to obtain the age of a single node $v_1$ exactly for any given network size $n$ and update rates $\lambda_e$ and $\lambda$. To find a closed-form \emph{order-wise} expression for the age in terms of the network size $n$, we can write upper and lower bounds for each recursion in \eqref{yates-fc}, in terms of the sum of reciprocal of integers, $\sum_{k=1}^i \frac{1}{k}$, and obtain the following upper and lower bounds for the age of a single node in the fully connected (FC) network,
\begin{align}\label{eqn: fc_age}
    \frac{\lambda_{e}}{\lambda}\left[\frac{n-1}{n}\sum_{k=1}^{n-1}\frac{1}{k} + \frac{1}{n}\right] \leq v_1^{FC} \leq \frac{\lambda_{e}}{\lambda}\sum_{k=1}^n \frac{1}{k},
\end{align}
from which we can conclude that the version age scaling for a single node in the fully connected network is $O(\log{n})$ as the sum of $\frac{1}{k}$ for $k$ from $1$ to $n$ grows as $\log n$.

\subsection{Age Scaling for a Bidirectional Ring Network}
The bidirectional ring network, as shown in Fig.~\ref{fig: rc_network}, is arranged in the form of a ring and each node communicates with its two neighbors, one on each side, as analyzed in \cite{yates21gossip} and \cite{buyukates22ClusterGossip}. The total rate of gossip for each node is $\lambda$, which it equally divides to $\frac{\lambda}{2}$ and $\frac{\lambda}{2}$ to gossip with each of its two neighbors. Similar to the case of fully connected network, here also, due to the symmetry of the network, the version age of a contiguous set of nodes depends only on the size of the set and not its position. We know that the version age of $\mathcal{N}$, $v_n = \frac{\lambda_e}{\lambda}$. Then, the recursion in \eqref{eqn: recursive equations}, expressed only in terms of the contiguous subset size $j$ is given by (see also \cite[Eqn.~(17)]{yates21gossip}),
\begin{align} \label{yates-br}
    v_j = \frac{\lambda_e + \lambda v_{j+1}}{\frac{j\lambda}{n} + \lambda}.
\end{align}
Again, in principle, one can start from $v_n = \frac{\lambda_e}{\lambda}$, and work backwards $j=n-1, n-2, \ldots, 2, 1$ utilizing \eqref{yates-br} to obtain the age of a single node $v_1$ in a bidirectional ring exactly. To find a closed-form \emph{order-wise} expression for the age in terms of the network size $n$, we can write upper and lower bounds for each recursion in \eqref{yates-br} in terms of sum of products terms. In this way, we find that the version age of a single node in the bidirectional ring network scales as follows,
\begin{align}\label{eqn: ring_age_apprx}
    v_1 \approx \frac{\lambda_{e}}{\lambda}\sum_{i=1}^{n-1}\prod_{j=1}^i\frac{1}{1+\frac{j}{n}}.
\end{align}
This can be written as an integral using a Riemann sum approximation via a step size of $\frac{1}{\sqrt{n}}$, yielding the result
\begin{align}\label{eqn: intg_apprx}
    \frac{1}{\sqrt{n}}\sum_{i=1}^{n-1}\prod_{j=1}^i \frac{1}{1+\frac{j}{n}} \approx \int_0^\infty e^{-\frac{t^2}{2}}dt = \sqrt{\frac{\pi}{2}},
\end{align}
from which we have $v_1 \approx \frac{\lambda_e}{\lambda}\sqrt{\frac{\pi}{2}}\sqrt{n}$. Thus, we can conclude that the version age scaling for a single node in a bidirectional ring network is $v_1=O(\sqrt{n})$.

Thus, the version age in fully connected network and ring network, which represent the two extremes of the connectivity spectrum, scales as $O(\log{n})$ and $O(\sqrt{n})$, respectively. Since both networks have the same update rates and consume similar bandwidth, we conclude that better connectivity leads to lower age, lower staleness, hence higher freshness, in the network. 

\begin{figure}[t]
    \centering\includegraphics[width = 0.85\linewidth]{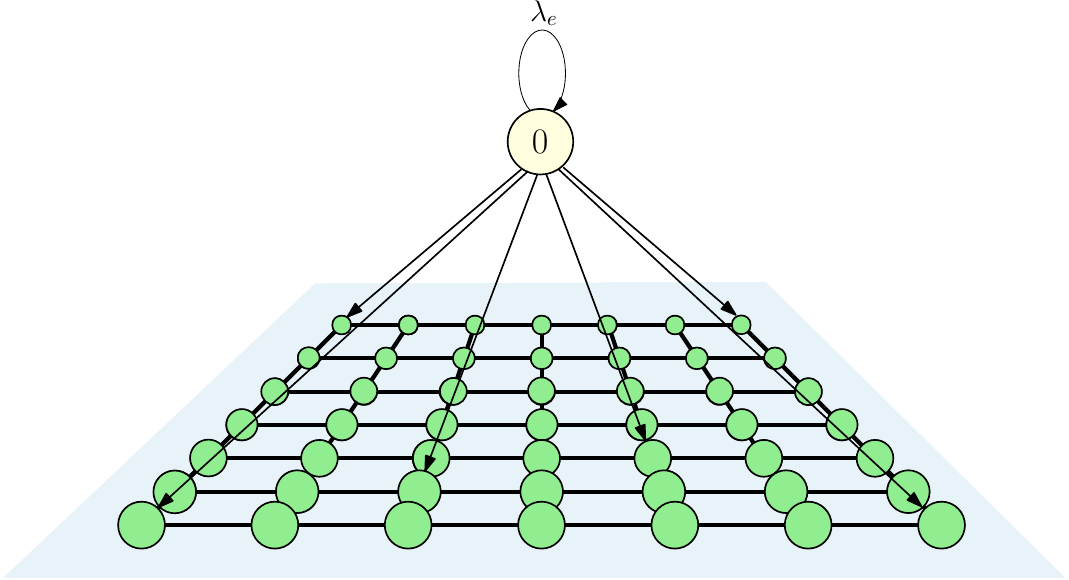}
    \caption{An illustration of the grid network.}
    \label{fig: grid network}
\end{figure}

\section{Age Upper Bound for Complex Networks}\label{sec: complex network}
In this section and the next, we analyze the version age scaling for more complex gossiping networks when compared to the bidirectional ring and the fully connected network. In the earlier cases studied so far, the number of recursive equations could be reduced from exponential to linear in the number of nodes in the network by exploiting the symmetry in the network; essentially, by using the fact that the age of a subset depends only on the size of the subset in those networks. This, however, is not possible for more complex networks, since arbitrary connections lead to many random connected sets, and age of a subset not only depends on the size of the subset but also on the specific shape of the subset, i.e., that no consolidation is possible only to the size in these cases.

Hence, in order to analyze version age scaling in such networks, we need to modify the recursive equations proposed in \cite{yates21gossip}, and use the geometry of these networks to find tight upper bounds. Specifically, we see that the version age of a set depends on the sets that are one-larger by including a neighbor of the set at each time. This evolution of the recursion is the reason for the exponential number of equations. Instead, we can write an upper bound for $v_S$ using only one-expanded set that has the highest version age among all one-expanded sets, and the number of neighbors of $S$, denoted by $N(S)$. This can also be modified to depend on the one-expanded set with the highest version age and the number of incoming edges (edges emanating at neighbors of $S$ and ending in a node in $S$) of set $S$. The method for bounding using $N(S)$ starts by rearranging \eqref{eqn: recursive equations} as follows,
\begin{align}\label{eqn: rearranged_recursion}
     \lambda_{00} = \lambda_0(S)v_S + \sum_{i \in N(S)}\lambda_i(S)(v_S - v_{S \cup \{i\}}).  
\end{align}
Next, we lower bound the sum on the right hand side as,
\begin{align}\label{eqn: upper_bound_technique}
    \!\!\!\lambda_{00} &\geq \lambda_0(S)v_S + |N(S)|\min_{i \in N(S)}\lambda_i(S)(v_S \!-\! v_{S \cup \{i\}})\\
    &\geq \lambda_0(S)v_S + |N(S)|\min_{i\in N(S)}\lambda_i(S) \min_{i \in N(S)}(v_S \!-\! v_{S \cup \{i\}}) \! \\
    &= \lambda_0(S)v_S + |N(S)|\min_{i \in N(S)}\lambda_i(S)(v_S \!-\! \max_{i \in N(S)} v_{S \cup \{i\}}). \! \label{eqn: eq7}
\end{align}
After rearranging the terms, we get an upper bound on $v_S$,
\begin{align}\label{eqn: age_upper_bound}
    \! v_S \leq \frac{\lambda_e + |N(S)| \min_{i\in N(S)}{\lambda_i(S)} \max_{i\in N(S)}{v_{S \cup \{i\}}}}{\lambda_0(S) + |N(S)|\min_{i\in N(S)}\lambda_i(S)}. \!
\end{align}

\begin{figure}[t]
    \centering\includegraphics[width = \linewidth]{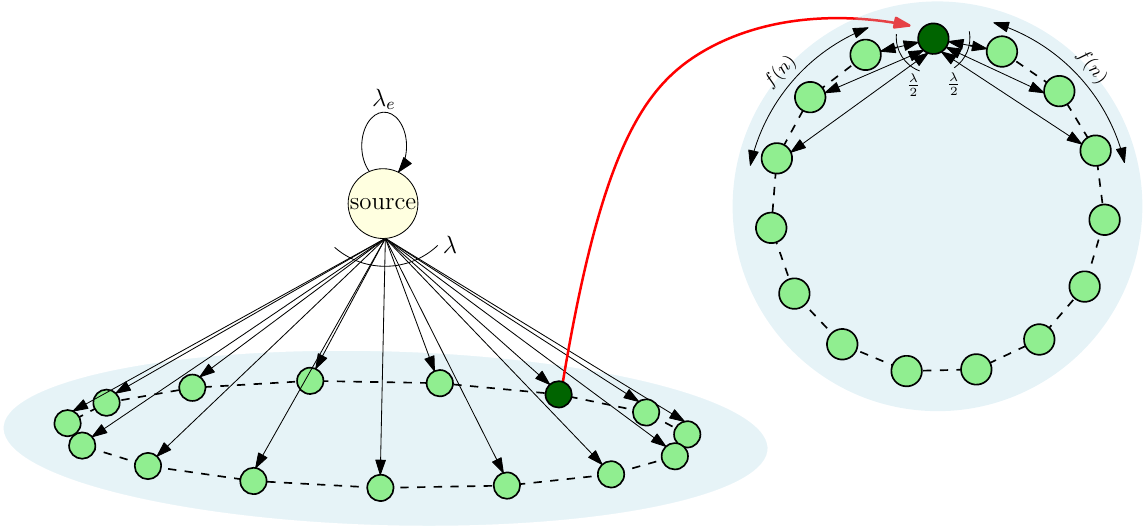}
    \caption{An illustration of the generalized ring network.}
    \label{fig: generalized ring network}
\end{figure}

The geometric parameters can be further lower bounded using the geometry of the network. The lower bound can depend on many parameters of the set. In order to reduce the number of equations from exponential to linear, we find the lower bound for the geometric parameters in terms of the size of the set. This lower bound may or may not result in a tight lower bound, depending on the generality of the networks. If the set of networks is too general, then the version age bounds are loose. This is because, in order to satisfy the geometric constraints for all networks in the set, we need to take into account sets with very bad connectivity. As an example, the set of all $d$-regular graphs \cite{Ellis2011TheEO} can be analyzed in this way, but the bounds do not provide insightful information about the real version age scaling of the graphs. On the other hand, we are able to  analyze two classes of networks: the two-dimensional grid network (Fig.~\ref{fig: grid network}) and the generalized ring network (Fig.~\ref{fig: generalized ring network}), and find tight upper bounds.

\subsection{Age Scaling for a Grid Network}\label{sec: grid}
\begin{figure}[t]
    \centering
    \includegraphics[width = 0.7\linewidth]{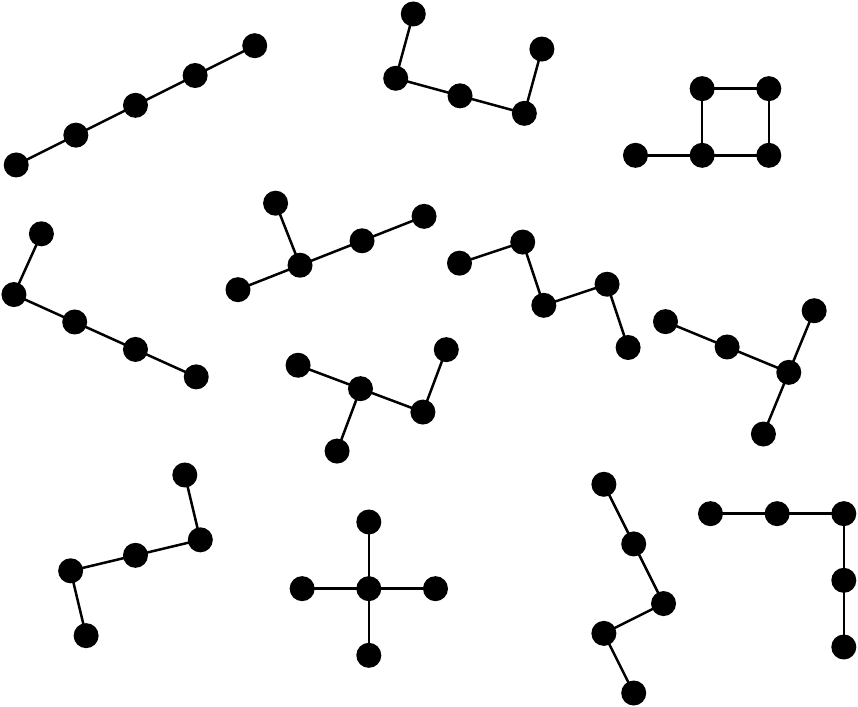}
    \caption{All subsets of the grid network containing five nodes.}
    \label{fig: 5sets}
\end{figure}

In \cite{srivastava2023grid}, we find the version age scaling for the two-dimensional grid network that has $n$ nodes, and hence is a $\sqrt{n} \times \sqrt{n}$ lattice. Fig.~\ref{fig: grid network} shows an illustration of the grid network. Unlike the ring and fully connected networks, the version age of a subset of the grid network depends not only on the size of the set, but also on its shape. The number of sets that have a fixed number of nodes increases rapidly. As an example, Fig.~\ref{fig: 5sets} shows all subsets of the grid network that contain $5$ nodes. Hence, directly applying the recursion in \eqref{eqn: recursive equations} is not feasible, since the number of equations grows rapidly as the size of the sets in the grid network increases. Instead, we use a lower bound on the number of incoming edges depending on the size of the set to find the upper bound for the recursion. In \cite{harary1976extremal}, it was found that the number of incoming edges of a set of $j$ nodes is lower bounded by $2\lceil 2\sqrt{j} \rceil$ in an infinite grid. In order to translate this result to our finite grid network, we need to consider the boundary effects. We see that the set with the least number of incoming edges changes as the number of nodes in the set increases. Hence, we derive a common lower bound to write the recursive upper bound equations. Then, we solve these equations and find that the version age scaling in the grid network is $O(n^{\frac{1}{3}})$. In comparison to the ring network, we see that the age scaling has improved significantly by adding only two extra connections per node. This is because the geometry of the grid network results in a $O(\sqrt{n})$ diameter of the network, whereas the diameter of the ring is $O(n)$. This facilitates faster transfer of information with gossiping.

\begin{figure}[t]
    \centering
    \includegraphics[width = 0.8\linewidth]{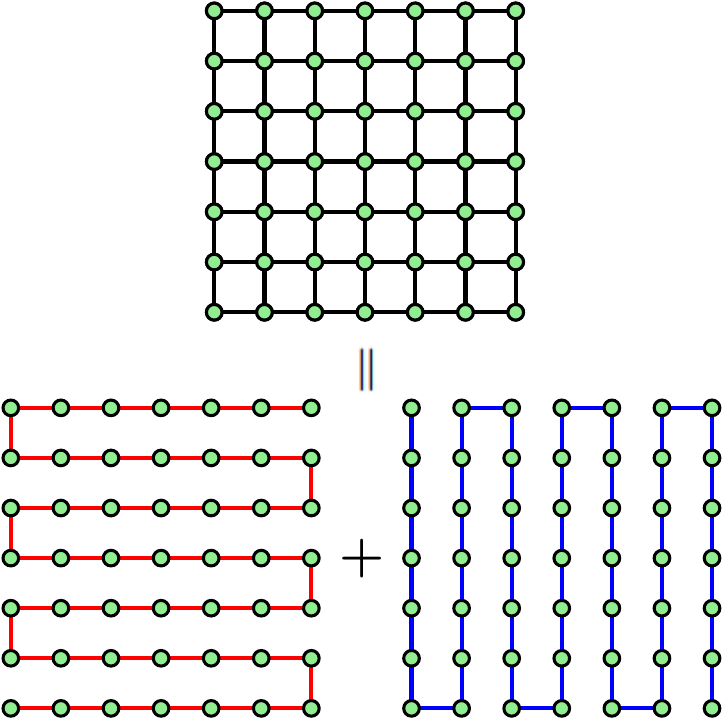}
    \caption{The grid can be thought of as two conjoined line networks, and a line network is equivalent to a ring network in terms of average version age \cite{kaswan_jamming_jrl}. Thus, conjoining two line networks to obtain a grid network reduces the version age from $O(n^{\frac{1}{2}})$ to $O(n^{\frac{1}{3}})$.}
    \label{fig: conjoined_grid}
\end{figure}

The grid network can also be considered from the perspective of conjoined networks. It is well known that conjoining two networks disseminating the same information speeds up the information diffusion \cite{yagan2013conjoining}. For example, suppose there is a group of friends living in the same city, who meet each other regularly. They want to be up-to-date about the happenings in each other's lives. We see that they will receive updates about their friends faster if they were connected to each other on a social media platform and also met each other in-person, rather than doing just one of the two. This is an interesting way with which we can also look at the grid network as a conjoined network of two line networks (the equivalence of version age scaling in line and ring networks was shown in \cite{kaswan_jamming_jrl}), as shown in Fig.~\ref{fig: conjoined_grid}, which improves the version age of the network from $O(n^{\frac{1}{2}})$ to $O(n^{\frac{1}{3}})$.    

\subsection{Age Scaling for a Generalized Ring Network}\label{sec: generalized_grid}
In \cite{srivastava2023generalizedrings}, we find the version age scaling for a network which is placed in a ring formation and each node now has $f(n)$ nodes on both sides, i.e., $2f(n)$ nodes in total, to gossip with. Fig.~\ref{fig: generalized ring network} shows an illustration of the generalized ring network. We analyze the network as $f(n)$ is varied, $1 \leq f(n) < n/2$. We use the number of incoming edges to write the recursive upper bound equations. It was shown in \cite{srivastava2023generalizedrings} that for sets of a fixed size with the least number of incoming edges is the set of contiguous nodes. Using this, lower bounds for the number of incoming edges were found for all values of $j$. The recursive upper bound equations are then solved to find the version age scaling of the generalized ring (GR) network as,
\begin{align}\label{eq: generalized ring}
    v_1^{GR} = O\left(\log{f(n)} + \sqrt{\frac{n}{f(n)}}\right).
\end{align}

We note that for two special cases studied already, i.e., when $f(n) = 1$, in which case the network becomes a ring network, and when $f(n) = n/2$, in which case the network becomes a fully connected network, \eqref{eq: generalized ring} reduces to their corresponding age expressions. In particular, \eqref{eq: generalized ring}  reduces to $O(\sqrt{n})$ in the first case, and to $O(\log{n})$ in the second case. Further, if $f(n)$ is any positive constant, which means that each node in the generalized ring network has a fixed number of neighboring nodes, then the version age still scales as $O(\sqrt{n})$ like a simple ring. Finally, if $f(n) = n^\alpha$, $0<\alpha<1$, i.e., $f(n)$ is a rational function, which covers a large range of functions between the two extremes of connectivity discussed in the first two cases, then the version age scales as $O(n^{\frac{1-\alpha}{2}})$. As a simple example, if $\alpha=\frac{1}{3}$, i.e., $f(n) = n^\frac{1}{3}$, then the age scales as $O(n^\frac{1}{3})$.

This work along with \cite{srivastava2023grid}, also allows us to analyze the dependence of the version age scaling for networks with a wide range of diameters and geometries. We note that the diameter of the generalized ring network is $O\left(\frac{n}{f(n)}\right)$. Hence, in order to have the same version age scaling as the grid network, we need $f(n) = n^{\frac{1}{3}}$. In other words, each node needs to be connected to $n^{\frac{1}{3}}$ nodes on each side in order to achieve the same version age scaling as the grid network, in which each node needs to be connected only to $4$ neighbors. From this, we can conclude that the version age scaling heavily depends on the geometry of the network.

The dependence of version age on the diameter of a generalized ring network can also be analyzed. In order to achieve a poly-logarithmic scaling in a generalized ring, we need the diameter to be only a poly-logarithmic gap away from $n$. 

When $f(n)$ is a rational function of $n$, i.e., $f(n) = n^\alpha$, $0<\alpha<1$, then the version age scaling of the network becomes $O(\alpha\log{n} + n^{\frac{1-\alpha}{2}})$. As $n \rightarrow \infty$, the second term, being a super-logarithmic function, dominates the first term. However, when $\alpha$ is large the second term grows very slowly, even slower than $\log{n}$. Hence, in order for us to say that the scaling is of the order of the second term, the number of nodes in the network need to be very large. In the range $0.6 \leq \alpha < 1$, we need the network to be in excess of a billion nodes for the version age scaling to be according to the second term. Since networks of this size do not occur in the real world, the version age scaling in this range for $\alpha$ can be considered to be logarithmic for all practical purposes.

\begin{figure}
    \centering
    \includegraphics[width = \linewidth]{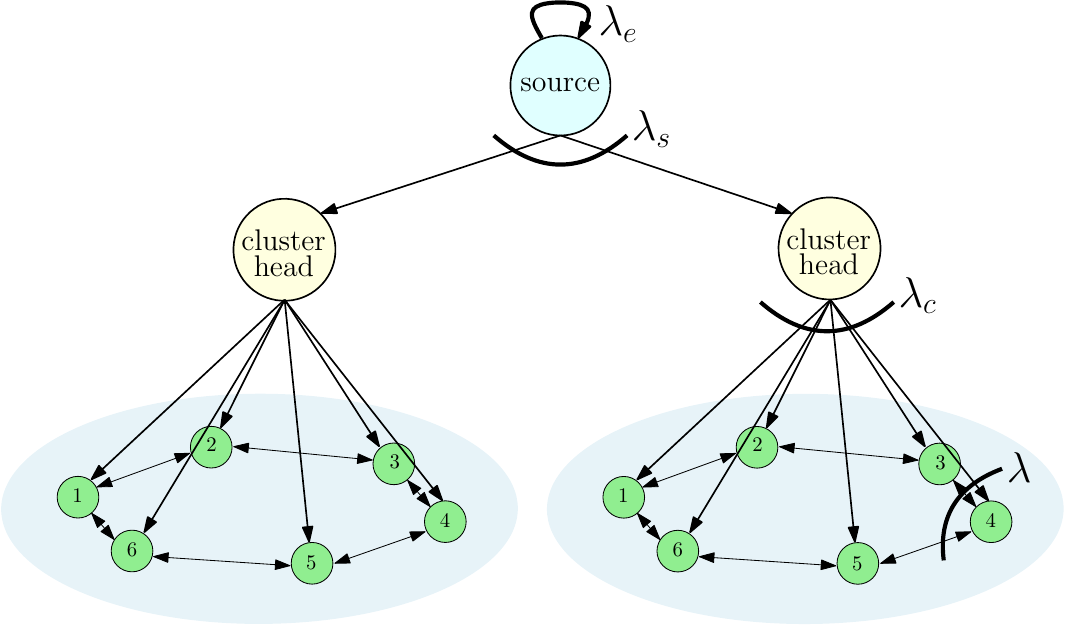}
    \caption{A gossip network with community (clustered) structure.}
    \label{fig: community structure}
\end{figure}

\section{Clustered Networks}\label{sec: clustered_network}
In this section, we describe networks that have hierarchical clustering and find the version age scaling. \cite{buyukates22ClusterGossip} investigated whether creating hierarchical networks can improve the version age scaling of nodes in the network. An example of this network is shown in Fig.~\ref{fig: community structure}. Clustered gossip networks have the usual source generating updates, cluster heads that get updated directly by the source and clusters consisting of nodes that are updated only by the respective cluster heads as a combined rate $\lambda_c$ Poisson process. Using a similar SHS calculation as in Section~\ref{sec: SHS_version_age}, we can obtain the following recursive equations,
\begin{align}\label{eqn: age_clustered}
    v_S = \frac{\lambda_e + \lambda_c(S)v_c + \sum_{i \in N_c(S)}\lambda_i(S)v_{S \cup i}}{\lambda_c(S) + \sum_{i \in N_c(S)}\lambda_i(S)}.
\end{align}

Along with these equations and the assumption that all the clusters have the same number of nodes and the same connectivity, we are able to find the version age scaling for several networks. There are two cases depending on whether the cluster heads themselves are connected to each other or not. If the nodes are connected to each other in a fully connected network, and the cluster heads are not connected to each other, then the version age scaling is still $O(\log{n})$. Hence, clustering does not improve the version age scaling for the highest possible connectivity beyond $\log n$; however, clustering reduces the connectivity requirements while achieving the same $\log n$ scaling (that is, nodes are fully connected only within the clusters, and disconnected across the clusters). On the other hand, for the ring and disconnected networks, the version age scaling improves from $O(n^{\frac{1}{2}})$ to $O(n^{\frac{1}{3}})$, and from $O(n)$ to $O(n^{\frac{1}{2}})$, respectively. Moreover, if the cluster heads are further connected in a ring network, then the version age scaling is further improved to $O(n^{\frac{1}{3}})$ in a disconnected cluster, and to $O(n^{\frac{1}{4}})$ in clustered ring networks. There is no improvement due to connection of cluster heads for clustered fully connected networks. Finally, if we have $h$ layers of hierarchy of rings, then the version age of each node scales as $O(n^{\frac{1}{2h}})$.

\begin{figure*}[t]
    \centering\includegraphics[width = 0.8\linewidth]{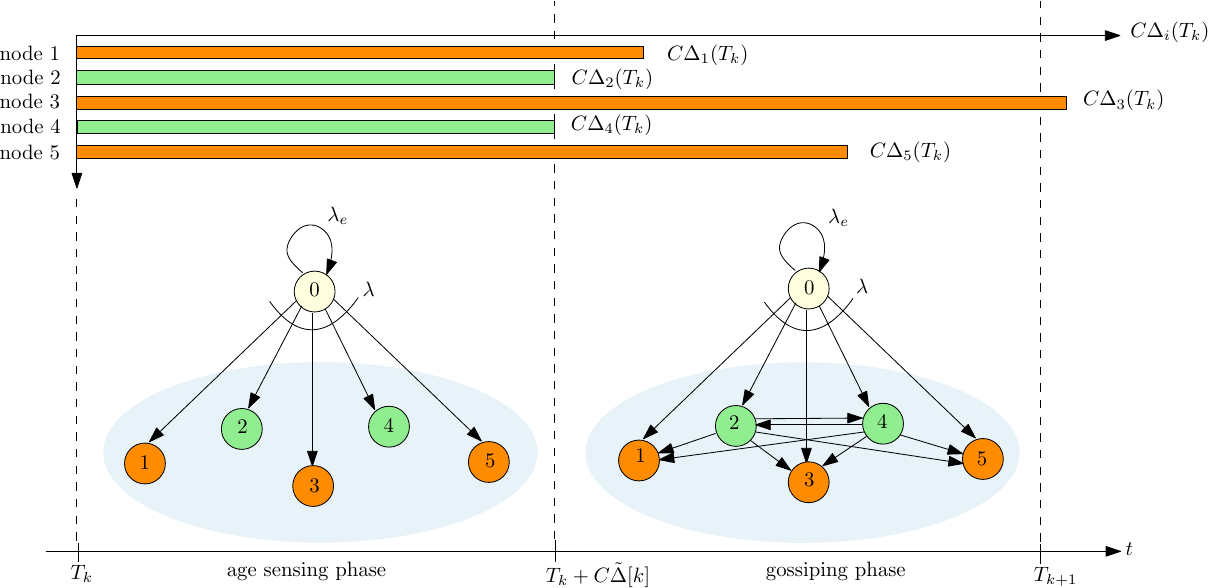} 
    \caption{Age-aware gossiping scheme ASUMAN.}
    \label{fig: asuman}
\end{figure*}

\section{Age-Aware Gossiping}\label{sec: age_aware_gossip}
In this section, we discuss a gossiping scheme for an age-aware, fully connected network. As discussed in Section~\ref{sec: age_scaling_simple-FC}, in a fully connected network, the average version age scales as $O(\log n)$. Even though this is better than the age scaling in all other lesser-connected networks (e.g., the ring network where age scaling is $O(\sqrt{n})$), it still grows as the network size grows. Here, we show that the age scaling can be brought down to $O(1)$, i.e., something that does not grow with the network size, if the nodes in the network are \emph{age-aware}, i.e., the nodes can estimate their own version ages. This age-awareness can be easily achieved by receiving a feedback signal from the source when the source updates itself. This version update information about the source, available at the network, can be leveraged for intelligent allocation of the overall gossip capacity. The uniform gossiping setting, described in the previous sections, where all network nodes have equal update capacity $\lambda$, allows nodes with relatively fresh information and nodes with relatively stale information to gossip with equal rate. Thus, a significant portion of the overall gossip capacity $B=n\lambda$ is wasted in communications that do not bring down the age of any node. Intuitively, if there is a mechanism such that the gossip capacity of the relatively higher age nodes can be shifted to the relatively fresh nodes, it should improve the overall timeliness of the network. The key challenge lies in implementing such a mechanism in a distributed manner, i.e., a mechanism where the nodes in the network will intelligently figure out the freshest nodes in the network at any particular time, and assign all the update capacities to them.

To that end, \textit{age sense update multiple access network (ASUMAN)} gossiping scheme was proposed in \cite{mitra_allerton22}. In this scheme, the age sensing nodes use the source self-update process as a method to synchronize themselves. When the source has a new version of information at time $T_k$, the nodes stop gossiping and wait for a time \emph{proportional to their own age}. This time interval is called the age sensing phase of the network. After this waiting time, each node sends a small pilot signal alerting their neighbors and start gossiping. The number of minimum age nodes can be estimated from the number of received pilot signals. Since the nodes with the minimum age at time $T_k$ have to wait for the least amount of time, only those nodes continue gossiping with rate $\frac{B}{|\mathcal{M}_k|}$, where $\mathcal{M}_k$ is the set of minimum age nodes at time $T_k$. On the other hand, the rest of the nodes $\mathcal{N}\backslash\mathcal{M}_k$ in the network remain silent up to time $T_{k+1}$; see Fig.~\ref{fig: asuman}.

The ASUMAN scheme yields an opportunistic mechanism, where the minimum age nodes are selected as the gossiping leaders. The average version age of the minimum age nodes, denoted as $\Tilde{X}(t)$, is independent of the network size $n$, and can be expressed as
\begin{align}\label{eqn: min_age_avg}
    \lim_{t\to\infty}\mathbb{E}\left[\Tilde{X}(t)\right]=\frac{\lambda_e+\lambda}{\lambda}.
\end{align}
Therefore, the ASUMAN scheme can be interpreted as the minimum age nodes, with average age obtained in \eqref{eqn: min_age_avg}, gossiping in the fully connected network with full capacity $n\lambda$. Hence, using SHS analysis, \cite{mitra_allerton22} shows that the average age of the $i$th node becomes
\begin{align}\label{eqn: age_ASUMAN}
    \lim_{t\to\infty}\mathbb{E}\left[X_i(t)\right]
    =\frac{\lambda_{e}}{\lambda}\frac{(1+\frac{n\lambda}{n-1}(\frac{1}{\lambda}+\frac{1}{\lambda_{e}}))}{\left(\frac{1}{n}+\frac{n}{n-1}\right)}\xrightarrow{ n \to \infty }2\frac{\lambda_e}{\lambda}+1.
\end{align}

The analysis in \cite{mitra2023age} shows that the ASUMAN scheme can be extended to networks with fractional connectivity, i.e., each node is connected to a fraction of $n-1$ nodes, or with networks where the source does not update all the nodes with equal rate. Additionally, using hierarchical structures with $O(n)$ connectivity among nodes can yield $O(1)$ age performance, although the upper bound is worse than that of the fully connected network. In this way, the trade-off between connectivity and age performance can be maintained. Importantly, \cite{mitra2023age} shows that ASUMAN does not produce good age performance for networks with finite $O(1)$ connectivity, such as ring, two-dimensional grid, etc., thus, sufficiently rich connectivity is needed to reap the benefits of opportunism.

The optimality of any gossiping scheme was shown in \cite{mitra_infocom23}, which proved a fundamental limit of age performance for gossiping with the total capacity of $B=n\lambda$ in a symmetric fully connected network. This is achieved by a semi-distributed scheme which lets the node with the minimum age at any time to gossip with full capacity. When the source updates a node, it becomes the minimum age node of the network. It sends a pilot signal in the network to alert the other nodes and starts gossiping. The gossiping continues until it receives a signal from any other node, which is the new minimum age node. The SHS analysis in \cite{mitra_infocom23} shows that this scheme is  optimal among all possible schemes with gossip capacity constraint $B=n\lambda$ and the achieved age performance is
\begin{align}\label{eqn: univ_lower_bound}
    \lim_{t\to\infty}\mathbb{E}\left[X_i(t)\right]=\frac{\lambda_e}{\lambda}\left(\frac{1+\frac{n}{n-1}}{\frac{1}{n}+\frac{n}{n-1}}\right)\xrightarrow{ n \to \infty }2\frac{\lambda_e}{\lambda}.
\end{align}

Additionally, \cite{mitra_infocom23} proposed a fully-distributed gossiping scheme that does not require any implicit coordination mechanisms, such as pilot signal transmissions in the network. In this scheme, the freshest node just gossips for a finite time duration of $\frac{1}{\lambda}$ and achieves an age performance of $(1+e)\frac{\lambda_e}{\lambda}$ for large $n$. We remark that although all these different schemes have different age performances, since they all scale as $O(1)$, the scaling results for different network topologies, such as fractional connectivity, finite connectivity, hierarchical connectivity, sublinear connectivity, derived for ASUMAN scheme in \cite{mitra_allerton22,mitra2023age} also apply to the semi- and fully-distributed schemes.
 
\section{Higher Order Moments of the Age Process}\label{sec: higher moments}
Reference \cite{abd2023distribution} considers the higher order moments of the age process. It uses the SHS method for evaluating the moment generating function (MGF) of a single node and the joint MGF of two age processes. For the subset of nodes $S, S_1, S_2\subseteq\mathcal{N}$, the test functions considered are $\psi^{(n)}_{S}(\pmb{X}(t))=\exp\left(nX_{S}(t)\right)$ and $\psi^{(n_1,n_2)}_{S_1,S_2}(\pmb{X}(t))=\exp\left(n_1X_{S_1}(t)+n_2X_{S_2}(t)\right)$.
By defining the rest of the reset maps and transitions similarly as before, \cite{abd2023distribution} shows that for any arbitrarily connected gossip network and for $m\geq 1$, the stationary marginal $m$th moment of the AoI for set $S\subseteq\mathcal{N}$ can be expressed as
\begin{align}\label{eqn: higher moments}
    {v}^{(m)}_{S}=\frac{m{v}^{(m-1)}_{S}+\sum_{i\in N(\mathcal{{S}})}\lambda_i(S){v}^{(m)}_{S\cup\{i\}}}{\lambda_0(S)+\sum_{i\in N(\mathcal{{S}})}\lambda_i(S)}.
\end{align}
Note for $m=1$, i.e., the first moment, \eqref{eqn: higher moments} reduces to \eqref{eqn: recursion_AoI}. 

\begin{figure}[t]
\centerline{\includegraphics[scale=0.6]{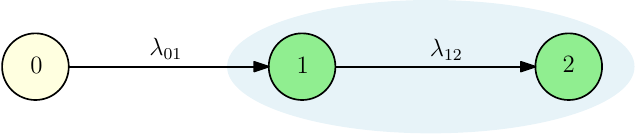}}
\caption{A toy example of gossip network for the calculation of higher order moments of the age processes.}
\label{fig: higher moments}
\end{figure}

Now, consider the simple toy example in Fig.~\ref{fig: higher moments}. Using \eqref{eqn: higher moments}, we obtain the variances of the two AoI processes, $X_1(t)$ and $X_2(t)$ at node $1$ and node $2$, respectively, as
\begin{align}\label{eqn: moments_toy_example}
    Var\left[X_1(t)\right]=\frac{1}{\lambda^2_{01}},\quad Var\left[X_2(t)\right]=\frac{1}{\lambda^2_{01}}+\frac{1}{\lambda^2_{12}}.
\end{align}
Thus, the results of \cite{abd2023distribution} show that the standard deviations of the age processes are relatively large with respect to their average (mean) values for gossip networks. Therefore, it is important to take higher order moments into consideration while implementing or optimizing timely gossip algorithms. 

\section{Sparse Networks}\label{sec: sparse network}
\begin{figure}[t]
\centerline{\includegraphics[scale=0.47]{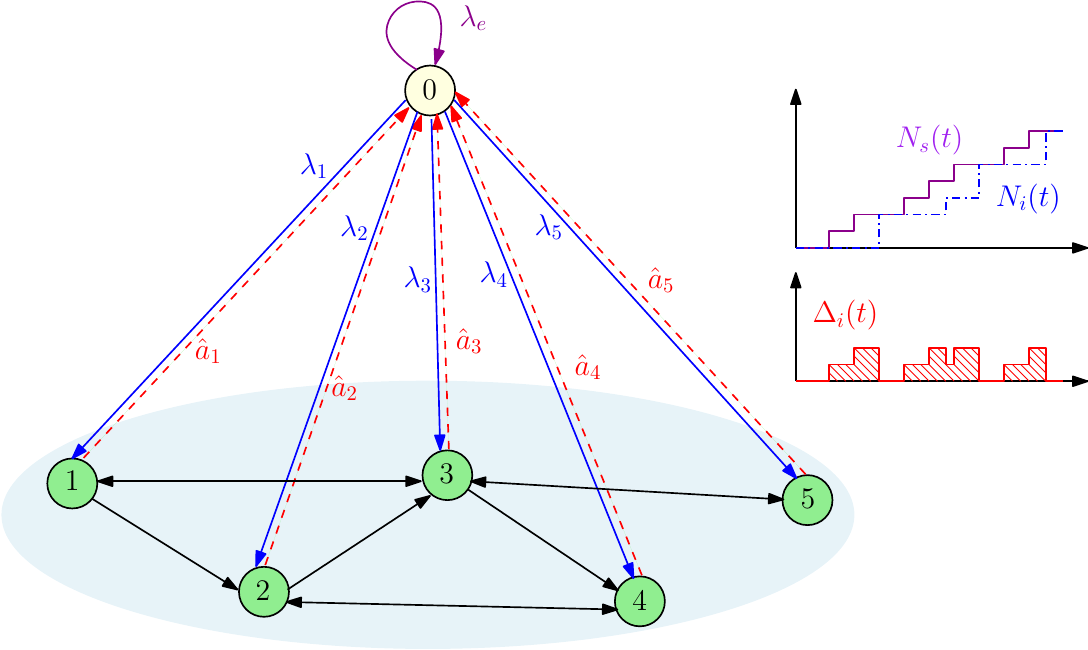}}
\caption{Learning based scheme for efficient allocation of update rates in a $5$-node sparse gossip network.}
\label{fig: learning update rate}
\end{figure}

The gossip networks we have discussed so far, are assumed to be symmetric, and hence easy to analyze. However, in real-life networks, such symmetric structures may not always be guaranteed; see Fig.~\ref{fig: learning update rate}. Several factors, such as link failures, expensive bandwidth, physical separation between devices, asymmetric connectivity, etc.~introduce heterogeneity in the network, which makes it difficult to analyze and optimize such networks. Indeed, applying the recursive formula in \eqref{eqn: recursive equations} with an asymmetric setting results in different age performance equations for different nodes, which are difficult to optimize with conventional gradient-based methods. 

To that end, \cite{mitra2023learning} introduces the concept of \emph{fair timeliness}, which evaluates the performance of the node with the worst average age in the network. The key idea behind this is that optimizing this worst case age performance will ensure a fair timeliness for all the other nodes in the network. The nodes are assumed to be age-aware, and therefore, they can calculate the empirical time average of their own ages as $\hat{a}_i=\frac{1}{T}\int_{T}\Delta_{i}(t)dt$. Assuming ergodicity of the age process, $\hat{a}_i\to a_i$ almost surely as $T\to\infty$, and therefore, for a large time window $T$, the estimate $\hat{a}_i\approx a_i$. Now, these average ages of the nodes can be tuned by choosing asymmetric update rates from the source to the node. Thus, the set of update rates from the source to the nodes $\boldsymbol{\lambda}=\{\lambda_i\}_{i=1}^{n}$ is the tunable parameter for the source; see Fig.~\ref{fig: learning update rate}. To maintain fair timeliness of the overall network, we minimize the average version age of the worst performing node $a(\boldsymbol{\lambda})=\max_{i\in\mathcal{N}}a_i(\boldsymbol{\lambda})$ by choosing $\boldsymbol{\lambda}$. Therefore, the optimization problem to solve is,
\begin{align}\label{eqn: original optimization}
    \min_{\boldsymbol{\lambda}\in[0,\lambda]^n} \ a(\boldsymbol{\lambda}) \qquad \text{subject to} \ \sum_{i=1}^{n}\lambda_i\leq\lambda.
\end{align}

Due to the recursive formulation of age in a sparse asymmetric network and the complications due to the $\max(\cdot)$ function, the solution of \eqref{eqn: original optimization} is obtained by a derivative-free or black-box optimization technique, i.e., a continuum multi-arm bandit formulation \cite{srinivas2009gaussian, bubeck2011x, shahriari2015taking, larson2019derivative} with action $\boldsymbol{\lambda}$ and reward $f(\boldsymbol{\lambda})=-\hat{a}(\boldsymbol{\lambda})$. Such formulation requires a sequential solution that makes the cumulative regret $R_M=\sum_{m=1}^{M}\left(\hat{a}(\boldsymbol{\lambda}_m)-\hat{a}(\boldsymbol{\lambda}^*)\right)$ sublinear, i.e., $o(M)$. \cite{mitra2023learning} uses sequential Bayesian optimization technique with Gaussian process regression, which fits a Gaussian process $GP(\mu(\boldsymbol{\lambda}),k(\boldsymbol{\lambda},\boldsymbol{\lambda}'))$ to $f$, where $\mu(\boldsymbol{\lambda})$ is the mean of the process and $k(\boldsymbol{\lambda},\boldsymbol{\lambda}')$ is a regularized kernel. Performing the regression for $M$ steps over the reward points $\boldsymbol{f}_M=[f_1, f_2, \cdots, f_M]^{T}$, and corresponding sequence $\{\boldsymbol{\lambda}_m\}$ yields the regression formula as
\begin{align}
    \mu_M(\boldsymbol{\lambda})& =\boldsymbol{k}_M(\boldsymbol{\lambda})^{T}\boldsymbol{K}_M^{-1}\boldsymbol{f}_M \label{eqn: mean regression}\\
    k_M(\boldsymbol{\lambda},\boldsymbol{\lambda}')& =k(\boldsymbol{\lambda},\boldsymbol{\lambda}')-\boldsymbol{k}_M(\boldsymbol{\lambda})^{T}\boldsymbol{K}_M^{-1}\boldsymbol{k}_M(\boldsymbol{\lambda}') \label{eqn: kernel computation}\\
    \sigma^2_M(\boldsymbol{\lambda})& =k_M(\boldsymbol{\lambda},\boldsymbol{\lambda}),\label{eqn: sigma regression}
\end{align}
where $\boldsymbol{k}_M(\boldsymbol{\lambda})=[k(\boldsymbol{\lambda},\boldsymbol{\lambda}_1),k(\boldsymbol{\lambda},\boldsymbol{\lambda}_2),\cdots,k(\boldsymbol{\lambda},\boldsymbol{\lambda}_M)]^{T}$ and $\boldsymbol{K}_M$ is the kernel matrix
\begin{align}\label{eqn: kernel matrix}
    \boldsymbol{K}_M\!=\!\!\!
    \begin{bmatrix}
    k(\boldsymbol{\lambda}_1,\boldsymbol{\lambda}_1)& k(\boldsymbol{\lambda}_1,\boldsymbol{\lambda}_2)& \cdots&k(\boldsymbol{\lambda}_1,\boldsymbol{\lambda}_M)\\
    k(\boldsymbol{\lambda}_2,\boldsymbol{\lambda}_1)& k(\boldsymbol{\lambda}_2,\boldsymbol{\lambda}_2)& \cdots& k(\boldsymbol{\lambda}_2,\boldsymbol{\lambda}_M)\\
    \vdots&\vdots&\vdots&\vdots\\
    k(\boldsymbol{\lambda}_M,\boldsymbol{\lambda}_1)& k(\boldsymbol{\lambda}_M,\boldsymbol{\lambda}_2)& \cdots&k(\boldsymbol{\lambda}_M,\boldsymbol{\lambda}_M)
    \end{bmatrix}.
\end{align}

Then, \cite{mitra2023learning} utilizes the commonly used upper confidence bound (GP-UCB) algorithm, which employs an optimization of a simple acquisition function over the search space. This auxiliary optimization can be written as
\begin{align}\label{eqn: acquisition function}
    \boldsymbol{\lambda}_m=\text{arg}\max_{\boldsymbol{\lambda}\in \mathcal{D}}\mu_{m-1}(\boldsymbol{\lambda})+\sqrt{\beta_m}\sigma_{m-1}(\boldsymbol{\lambda}),
\end{align}
where $\mathcal{D}=\{\boldsymbol{\lambda}\in[0,\lambda]^n:\mathbbm{1}^T\boldsymbol{\lambda}\leq \lambda\}$ is the search space. This optimization is much simpler than the original optimization and can be performed with the existing gradient-based numerical methods. The term $\mu_{m-1}(\boldsymbol{\lambda})$ in \eqref{eqn: acquisition function} exploits information from the data points up to $m-1$ steps. Whereas the term $\sigma_{m-1}(\boldsymbol{\lambda})$ pushes the algorithm for exploring different regions of the search space, thus meeting the exploration-exploitation trade-off. Analysis in \cite{srinivas2009gaussian} shows that choosing $\beta_m\sim O(\log(m^2))$ for a convex search space, such as $\mathcal{D}$, the regret $R_M$ is guaranteed to be sublinear with arbitrary high probability, i.e., with high probability the algorithm converges to a global optimum. The only downside of using GP-UCB is that the auxiliary optimization is not scalable, as the time complexity increases exponentially for large networks.

\section{File Slicing and Network Coding}\label{sec: file_slicing}
In Section~\ref{sec: age of gossip}, we saw how in a complete symmetric network of $n$ nodes that wishes to track a single time-varying file, the average age at each node comes out to $O(\log n)$. A natural question that follows is whether we can do better than $O(\log n)$ version age in a fully connected network.\footnote{We remark that the work summarized in Section~\ref{sec: age_aware_gossip} achieves this goal by utilizing two modifications to the system model: Nodes are age-aware and the total network gossiping capacity can be redistributed. In the current section, via use of file slicing and periodic pausing, we achieve $O(1)$ age while being age-agnostic and without redistributing gossiping capacity.} Likewise, if $n$ time-varying files, all generated at distinct nodes, were simultaneously tracked by all $n$ nodes, then dividing all rate resources among $n$ files in \cite{yates21gossip} results in a version age of order $O(n \log n)$ with respect to each file. Therefore, one can ask whether we can do better than $O(n \log n)$ for $n$ files. In this respect, \cite{kaswan22slicingcoding} achieves $O(1)$ version age for single-file dissemination and $O(n)$ version age for $n$-file dissemination (improving both results by $\log n$), in a network of $n$ nodes, though their model is slightly different from \cite{yates21gossip}.

\begin{figure}[t]
    \begin{center}
    \subfigure[]{\includegraphics[width=0.49\linewidth]{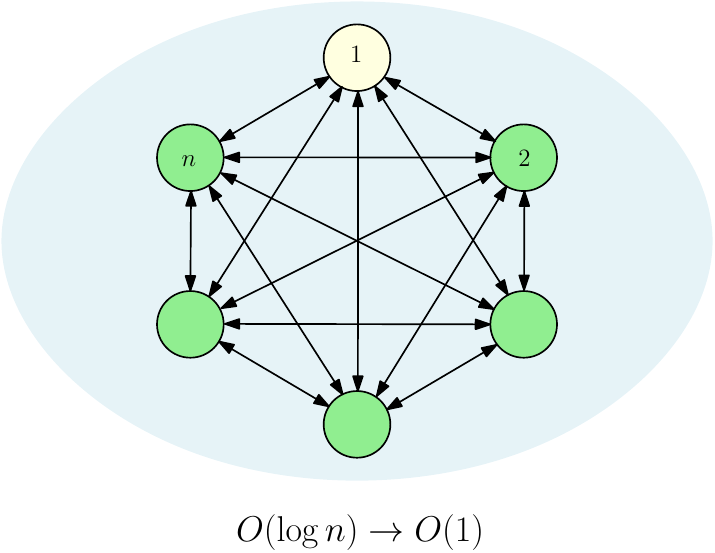}}
    \subfigure[]{\includegraphics[width=0.49\linewidth]{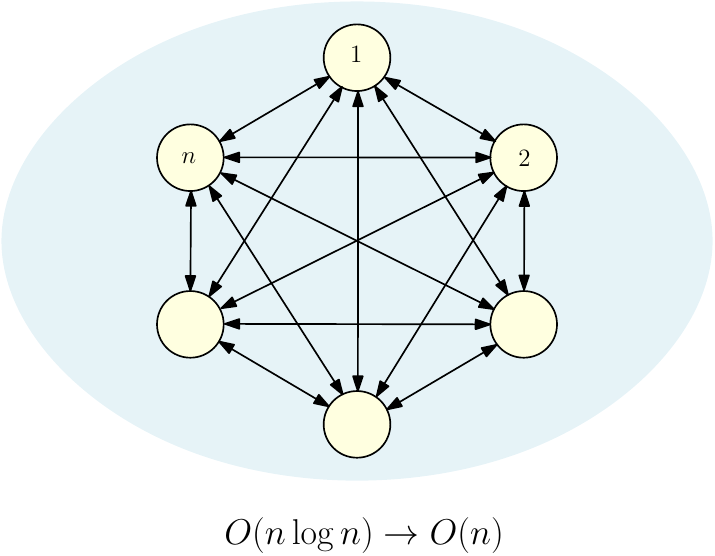}}
    \end{center}
    \caption{Fully connected network of $n$ nodes with (a) node $1$ as source node of single file, (b) all nodes as source nodes of distinct file.}
    \label{fig:slicing_coding}
 \end{figure}

\cite{kaswan22slicingcoding} assumes a discrete time model of gossiping, where the timeline is divided into cycles, with each cycle containing $O(1)=c$ timeslots in the case of single file dissemination; see Fig.~\ref{fig:slicing_coding}(a). Here one timeslot is the duration of time required to receive an entire file at a node. The constant $c$ is determined based on the specifics of the gossip protocol, since their scheme works for a wide class of gossip protocols that satisfy certain conditions. At the beginning of each new cycle, the source begins transmitting the latest version in its possession to other nodes, and the latter further transmit packets to neighbors as part of gossiping. The age analysis depends on one key factor -- in a single cycle, a fixed version is gossiped in the network to avoid mixing of versions. That is, at the beginning of a new cycle, all nodes halt transmissions of previous versions they obtained in prior cycles and the source further does not transmit any new versions that it gets updated with during the course of the current cycle. 

\cite{kaswan22slicingcoding} employs a class of gossip protocols that have a non-zero probability of transmitting a fixed file to all network nodes in $O(1)=c$ time. \cite{kaswan22slicingcoding} shows how the INTERLEAVE protocol of \cite{Sanghavi2007GossipFileSplit} mentioned in Section~\ref{sec: intro} can be fine-tuned to satisfy such requirements, which slices packets into smaller pieces and uses a push-pull hybrid scheme to disseminate the sliced packets. The effectiveness of the periodic pausing of newer version updates in each cycle in achieving expected age of $O(1)$ is proven by graphically studying the age at an arbitrary node within a typical cycle and formulating an upper bound that is $O(1)$ in expected sense.

The analysis can be extended to the case of multi-file gossiping shown in Fig.~\ref{fig:slicing_coding}(b), where the system consists of a network of $n$ nodes and $n$ files, such that each node is the source of a unique time-varying file that all other nodes aim to closely track. Again the timeline is divided into cycles, such that each cycle consists of $cn$ timeslots for some constant $c$ chosen based on specifics of the gossip protocol. In this case, a class of gossip protocols capable of disseminating a fixed set of $n$ messages to all nodes within $cn$ timeslots in each cycle is employed; an example of such a gossip protocol is the random-linear coding (RLC)-based protocol discussed in \cite{deb2006AlgebraicGossip}. In each cycle, this gossip protocol is utilized to distribute the latest versions of all $n$ files possessed by all source nodes at the beginning of the cycle. Through the design of a pertinent upperbound, it is demonstrated that the age for each file at each node is $O(n)$.

In both the cases of single file and $n$ files dissemination, two noteworthy aspects emerge. Firstly, achieving the specified age bounds does not necessitate the complete dissemination of a file version to the entire network in each cycle. Even if very few nodes really receive a file version in a particular cycle, since every cycle presents a fresh opportunity for a node to receive a new version and reduce its age, it is okay for network nodes to miss out some updates in a couple of cycles. Further, periodic pausing of dissemination of newer file versions from the source does not hinder the achievement of these bounds.

\section{Reliable and Unreliable Sources}\label{sec: reliable_unreliable}
\begin{figure}[t]
\centerline{\includegraphics[width=0.7\linewidth]{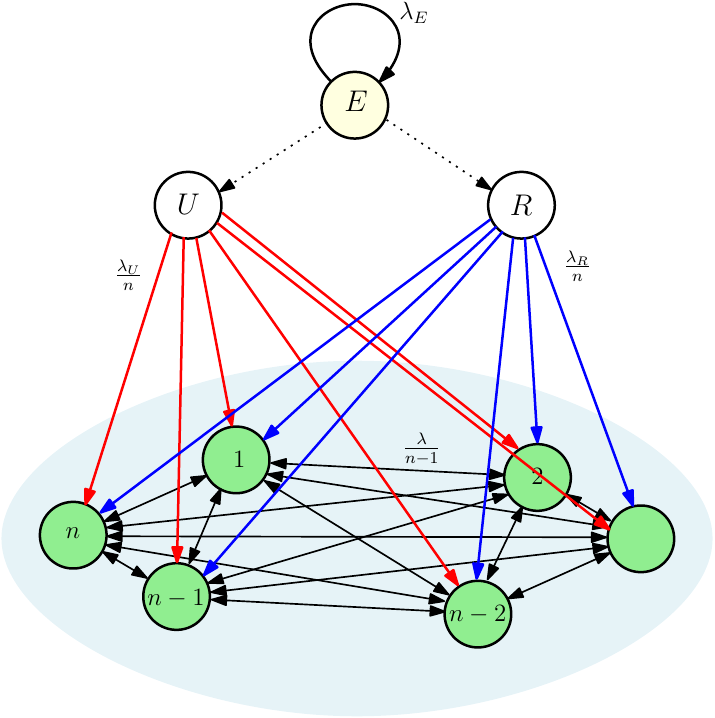}}
\caption{System model with reliable ($R$) and unreliable ($U$) sources observing an event ($E$) that updates with rate $\lambda_E$. Reliable ($R$) and unreliable ($U$) sources disseminate the information to the network at total rates $\lambda_R$ and $\lambda_U$, respectively. Each user gossips with its neighbors with rate $\lambda$ over a fully connected network.}
\label{fig: system_modelRU}
\end{figure}

In the works so far, we have examined system models involving a source node receiving updates for a time-varying file and accurately disseminating updates on the current status of the file to the network. The work in \cite{Kaswan23reliable_Ggap} delves into the system model depicted in Fig.~\ref{fig: system_modelRU}. In this model, a network comprising $n$ nodes, denoted as $\mathcal{N}=\{1,\ldots,n\}$, aims to continually track an updating process or event ($E$) that receives version updates according to a Poisson process with a rate of $\lambda_E$. However, for transmitting information about the event to the nodes, two sources are available —- a reliable source and an unreliable source. Information received from the former is considered more reliable than that from the latter and is expected to provide more accurate information. The less reliable source could, for instance, act as a proxy for a cost-effective sensor transmitting quantized or noisy measurements to an IoT network. This system model distinguishes itself from previous works in that there is still a single source of information, but two relaying sources are now available.

Nodes wish to have fresh information, however, they have preference for packets that originated at the reliable source, i.e., reliable information, and are willing to sacrifice their version age of information by up to $G$ versions to switch from an unreliable packet to a reliable packet. Consider $S_i(t)$ as an indicator of the reliability status of the information packet at node $i$ at time $t$, where $S_i(t)=0$ and $S_i(t)=1$ represent reliable and unreliable packets, respectively. Additionally, let $X_i(t)$ denote the version age of information at node $i$ at time $t$. At time $t$, when node $i$ sends an update to node $j$, node $j$ decides to accept or reject the packet based on the following rules: If both nodes have unreliable information, node $j$ selects the packet with the lower version age. If both nodes have reliable information, node $j$ chooses the packet with the lower version age. If the incoming packet has reliable information but node $j$ has unreliable information, node $j$ selects the incoming reliable packet if $X_i \leq X_j+G$. If node $j$ has a reliable packet and the incoming packet is unreliable, node $j$ retains its reliable packet if $X_j \leq X_i+G$. Let $F(t)= \frac{S_1(t)+ S_2(t)+ \ldots + S_n(t)}{n}$ denote the fraction of user nodes that have unreliable information packet at time $t$.

The goal is to study how this protocol impacts the prevalence of unreliable packets at nodes in the network and their version age. Using the SHS framework, \cite{Kaswan23reliable_Ggap} formulates analytical equations to characterize two quantities: long-term expected fraction of nodes with unreliable packets $F= \lim_{t \to \infty} \mathbb{E}[F(t)]$ and expected version age of information at network nodes $x_i= \lim_{t \to \infty} \mathbb{E}[X_i(t)]$, $i \in \mathcal{N}$. The choice of the continuous state for the SHS is $(\pmb{S}(t),\pmb{X}(t))\in \mathbb{R}^{2n}$, where $\pmb{S}(t)=[S_1(t),\ldots,S_n(t)]$ and $\pmb{X}(t)=[X_1(t),\ldots,X_n(t)]$ represent the instantaneous reliability status and instantaneous version age, respectively, of the $n$ user nodes at time $t$. The SHS operates in a single discrete mode with the differential equation $(\pmb{\dot S}(t),\pmb{\dot X}(t))=\pmb{0}_{2n}$. 

For a set of nodes $A$, \cite{Kaswan23reliable_Ggap} introduces the concepts of reliability status $S_A$ and version age $X_A$, which play a crucial role in assessing certain test functions later. The determination of $S_A$ and $X_A$ essentially involves identifying the most optimal node in the set $A$ in some sense, such that its reliability status and version age become representative of the entire set. If the most recent reliable packet is at most $G$ versions older than the latest unreliable packet in the node set $A$, then the node with the latest reliable packet establishes the values of $X_A$ and $S_A$. Otherwise, if the most recent unreliable packet is predominant, the node with the latest unreliable packet determines $X_A$ and $S_A$. With these definitions, \cite{Kaswan23reliable_Ggap} introduces five distinct test functions, leading to five separate recursive equations, unlike the single test function and recursive equation (\ref{eqn: recursive equations}) in classical age-based gossiping discussed in Section~\ref{sec: age of gossip}. These five equations can be solved iteratively to obtain values for $F$ and $x_i$. \cite{Kaswan23reliable_Ggap} then proves several results that allow us to show that $F$ is a decreasing function of $G$ and $x_i$ are an increasing function of $G$. Therefore, $G$ induces a trade-off between reliability and freshness of information.

\begin{figure}[t]
\centerline{\includegraphics[width=0.75\columnwidth]{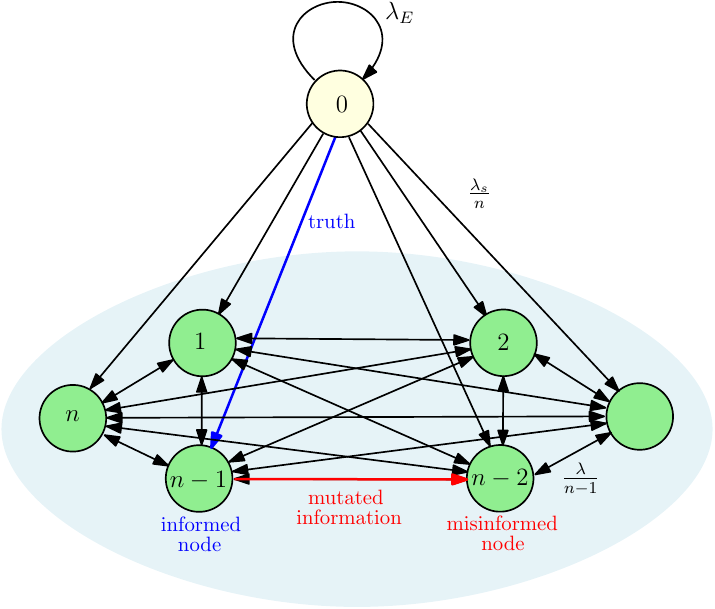}}
\caption{Fully connected gossip network of $n$ nodes with information probabilistically mutating into misinformation during inter-node gossip.}
\label{fig: system_model_mutation}
\end{figure}

\section{Information Mutation}\label{sec: info_mutation}
We proceed to extend the model depicted in Fig.~\ref{fig: fc_network} to account for alterations in information as it passes from one node to another; see Fig.~\ref{fig: system_model_mutation}. \cite{kaswan23mutation} attempts to characterize spread of misinformation in an age-based gossip network of $n$ user nodes that receives version updates from a source. The source always communicates accurate information to network nodes, however, there is a possibility of information getting mutated during inter-node transmissions in the network. This can stem from the sender node not always being truthful, occasionally deliberately altering the information before forwarding it, or the information packet becoming corrupted during the transmission process. An exemplar of this scenario could be software distribution to end users, where the software vendor consistently provides reliable versions or iterations of software packages to users. However, if users acquire a software version from their neighboring users, they might sporadically receive an incorrectly functioning or even harmful version of the software.

Here, $T_i(t)$ represents the accuracy of information at node $i$, where $T_i(t)=1$ indicates accurate information (as originated at the source, also termed the truth), and $T_i(t)=0$ indicates inaccurate information (alternatively referred to as misinformation). Upon receiving a packet, node $i$ cannot immediately discern whether the information is true or not, or if it differs from the sender's information. Upon receiving a packet, node $i$ compares its version number with the received packet; if different, the staler packet is discarded, and the fresher one is retained. However, if the version numbers match, node $i$ retains the information it trusts more, perhaps based on software performance or measurement noise in a smart sensor network. The assumption, then, is that when two packets have the same version age, truth prevails over misinformation; thus, if one of the packets contains accurate information (the truth), node $i$ retains that packet.

The goal here is to find what fraction of nodes are misinformed in this network and the version age of information at all user nodes, which is somewhat similar to \cite{Kaswan23reliable_Ggap}. However, a distinction lies in \cite{Kaswan23reliable_Ggap} where once a packet is created at one of the sources, the packet information remains unchanged during the diffusion process. In contrast, in \cite{kaswan23mutation}, information is susceptible to mutation into misinformation during inter-node transmissions. Additionally, in \cite{Kaswan23reliable_Ggap}, network nodes are aware of whether a particular packet originated at the reliable or unreliable source, allowing them to consider a freshness-reliability trade-off. Conversely, in \cite{kaswan23mutation}, nodes are unaware of whether the received information is the truth or not.

The information mutation problem is addressed through SHS modeling, where the continuous state is denoted as \((\pmb{T}(t),\pmb{X}(t))\in \mathbb{R}^{2n}\), with \(\pmb{T}(t)=[T_1(t),\ldots,T_n(t)]\) and \(\pmb{X}(t)=[X_1(t),\ldots,X_n(t)]\) representing the instantaneous accuracy and instantaneous version age, respectively, of packets stored at \(n\) user nodes at time \(t\). A transition \((i,j,h)\) occurs when node \(i\) sends a packet to node \(j\), where \(h=1\) signifies error-free communication, and \(h=0\) indicates mutation of information into misleading information during transmission. Transition \((0,0,1)\) represents an update at the source. The stochastic hybrid system operates in a single discrete mode, with the continuous state following the differential equation \((\pmb{\dot T}(t),\pmb{\dot X}(t))=\pmb{0}_{2n}\). In \cite{kaswan23mutation}, variables \(T_{A,B}\) are defined for two disjoint sets of nodes \(A\) and \(B\), leading to the introduction of three test functions and corresponding recursive equations. These equations must be iterated in a specific manner to solve for the desired quantities: the expected fraction of users with truth \(F= \lim_{t \to \infty} \mathbb{E}[F(t)]\) and version age \(x_i= \lim_{t \to \infty} \mathbb{E}[X_i(t)]\), \(i \in \mathcal{N}\), where \(F(t)= \frac{T_1(t)+ T_2(t)+ \ldots + T_n(t)}{n}\).

Noteworthy findings in \cite{kaswan23mutation} include the observation that very low gossip rates control the dissemination of mutated information on one hand, while very high gossip rates accelerate the dissemination of accurate fresh packets to all network nodes, mitigating misinformation due to the prevalence of truth in the model on the other hand. Hence, both extreme cases contribute to minimizing misinformation, with misinformation spread being higher at moderate gossip rates.

\begin{figure}[t]
\begin{center}
    \subfigure[]{\includegraphics[width=0.49\linewidth]{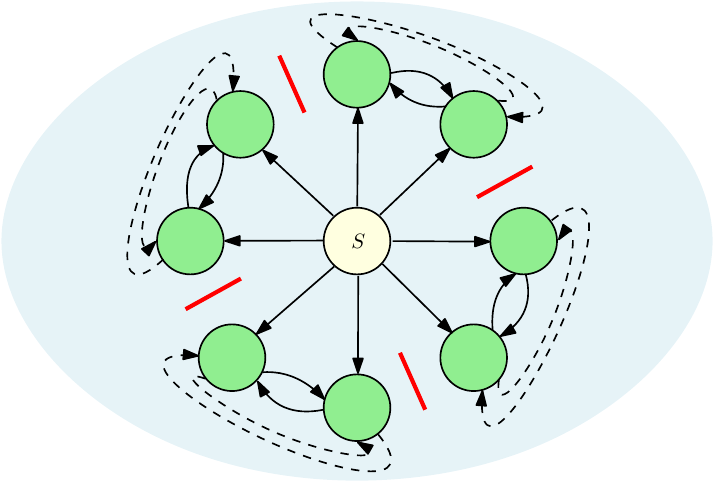}}
    \subfigure[]{\includegraphics[width=0.49\linewidth]{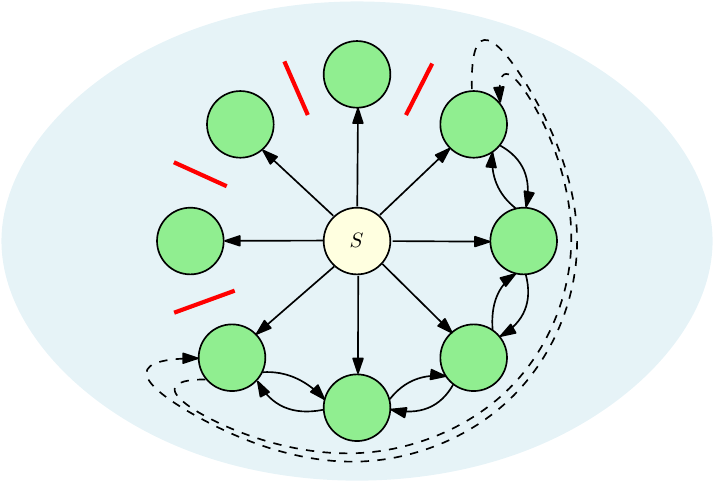}}
    \end{center}
\caption{Ring network with source node (yellow) and $n$ user nodes (green) with $\tilde n$ jammed or missing links (red lines). Addition of hypothetical links (dotted lines) transforms each individual line of nodes into a small ring. (a) Least harmful jammer placement: equidistant placement. (b) Most harmful jammer placement: adjacent placement.}
\label{fig:ring_network_jammed}
\end{figure}

\section{Adversarial Actions in Gossiping}\label{sec: adversary}
Up until this point, our exploration has focused on information distortion in gossiping in the form of either information dissemination by a less reliable source in \cite{Kaswan23reliable_Ggap}, or innocent mutations to information packets happening during transmission in \cite{kaswan23mutation}. However, these works did not involve a malicious entity actively attempting to disrupt the gossip operation. In this section, we will delve into and review two studies that explore adversarial attacks —- specifically, jamming attacks and timestomping attacks.

\subsection{Jamming Attack}\label{sec: jamming}
\cite{kaswan_jamming_jrl} investigates the resilience of age-based gossiping over networks against intentional jamming. It focuses on characterizing the impact of the number of jammers $\Tilde{n}$ on the version age scaling in two distinct gossip network topologies, the ring network and the fully connected network, which represent the two extremes of the connectivity spectrum of networks. In the ring network, it is shown that when the number of jammers $\tilde{n}$ scales as a fractional power of network size $n$, i.e., $\tilde n= cn^\alpha$, the average version age scales with a lower bound $\Omega(\sqrt{n})$ and an upper bound $O(\sqrt{n})$ when $\alpha \in \left[0,\frac{1}{2}) \right.$, and with a lower bound $\Omega(n^{\alpha})$ and an upper bound $O(n^{\alpha})$ when $\alpha \in \left[\frac{1}{2},1\right]$. This implies that the version age with gossiping on a ring remains robust against upto $\sqrt{n}$ jammers, considering that the version age in a ring network without jammers scales as $\sqrt{n}$. These age scalings hold irrespective of the particular placement of jammers on the gossiping links. 

When multiple adversaries cut inter-node communication links in this symmetric ring, the ring network is dismembered into a collection of isolated groups of nodes as in Fig.~\ref{fig:ring_network_jammed}, where each group has the structure of a \emph{line network}. The age of nodes in each such group are no longer statistically identical, owing to disappearance of circular symmetry. The spatial variation of version age over a line network is initially examined, where it is shown that expected version age in this network is highest at the corner nodes and decreases towards the center. Subsequently, bounds on version age of line network, and, consequently, on the dismembered ring are derived. For this purpose, \cite{kaswan_jamming_jrl} introduces an alternate system model of mini-rings (formed by dotted lines, as depicted in Fig.~\ref{fig:ring_network_jammed}, around each individual line network component, thus forming a smaller ring with fewer than $n$ nodes) and proves that the version age of the original model can be sandwiched between constant multiples of the version age of the alternate mini-ring model. Several other structural results are also proven along the way.  

\begin{figure}[t]
\centerline{\includegraphics[width=1\linewidth]{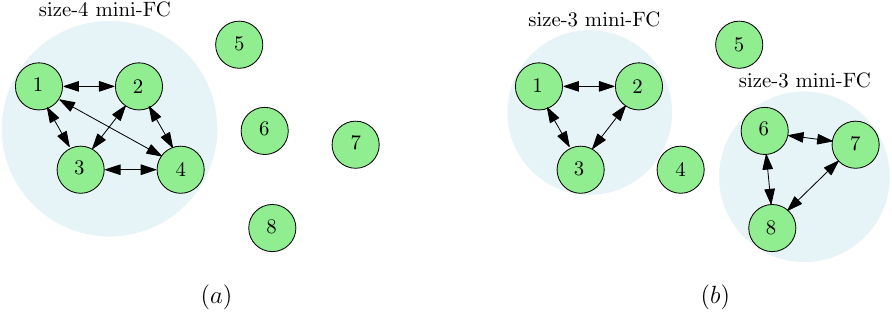}}
\caption{A network of $n=8$ nodes with $\binom{n}{2}-\Tilde{n}=6$ remaining links. The network configuration of (a) has the $6$ links consolidated, thereby isolating $4$ nodes from gossiping, which results in higher age than configuration of (b) where the links are more evenly distributed amongst nodes.}
\label{fig:clustering_miniFC}
\end{figure}

In the case of the fully connected gossip network, \cite{kaswan_jamming_jrl} formulates a greedy strategy to place $\Tilde{n}$ jammers so as to maximize the age of the resultant network. The greedy method involves using the $\Tilde{n}$ jammers to isolate maximum possible nodes, thereby consolidating all the remaining links into a single ball, as shown in Fig.~\ref{fig:clustering_miniFC}(a). In the resulting network, the average version age is shown to scale as $O(\log{n})$ when $\Tilde{n}=O(n\log{n})$, and as $O(n^{\alpha-1})$, $1<\alpha\leq2$ when $\Tilde{n}=O(n^{\alpha})$, implying that the network is robust against $n\log{n}$ jammers, since the version age in a fully connected network without jammers scales as $O(\log{n})$. Therefore, despite having the same update capacity in both ring and fully connected networks, the large number of links in the fully connected topology constrains the effectiveness of jammers, requiring higher number of jammers for any meaningful deterioration of system age scaling of the network.

\subsection{Timestomping Attack}\label{sec: timestomping}
\cite{kaswan_timestomp_jrl} explores timestomping attacks in timely gossiping, where an adversary disrupts the gossip operation by manipulating timestamps of certain data packets in the network. This technique, referred to as \emph{timestomping} \cite{minnaard2014timestomping}, aims to introduce staleness and inefficiency into the network. A timestomping attack can take various forms. A malicious insider node may breach the gossip protocol, injecting old packets by disguising them as fresh through timestamp manipulation, while maintaining the gossiping frequency to evade suspicion. Alternative methods include \emph{meddler in the middle} (MITM) attacks, where the adversary inserts its node undetected between two nodes and manipulates communication, and \emph{eclipse} attacks where the adversary redirects a target node's inbound and outbound links to adversary-controlled nodes for manipulation, isolating it from the legitimate network.

Consider two nodes, $A$ and $B$, coming in contact to exchange information, such that node $A$ is controlled by a timestomping adversary. If node $A$ is outdated compared to node $B$, the adversary is inclined to increase the timestamp of an outgoing packet from node $A$ to make it appear fresher so as to misguide node $B$ into discarding its packet in favor of a staler packet, and also, decrease the timestamp of an incoming packet from node $B$ so as to avoid its acceptance at node $A$. Conversely, if node $A$ is more up-to-date than node $B$, the adversary would reduce timestamps of outgoing packets and increase timestamps of incoming packets to make node $B$ reject fresher files and node $A$ accept staler files. The further the manipulated timestamps deviate from their actual values, higher are the chances of error in deciding which packet should be discarded, since this decision relies on a comparison of timestamps. At time $t$, the maximum error occurs when the timestamp is changed to either the current time $t$ or the earliest time $0$. In this respect, \cite{kaswan_timestomp_jrl} investigates an oblivious adversary that probabilistically alters the timestamp of each incoming and outgoing packet to either $t$ or $0$.

\begin{figure}[t]
    \begin{center}
    \subfigure[]{\includegraphics[width=0.49\linewidth]{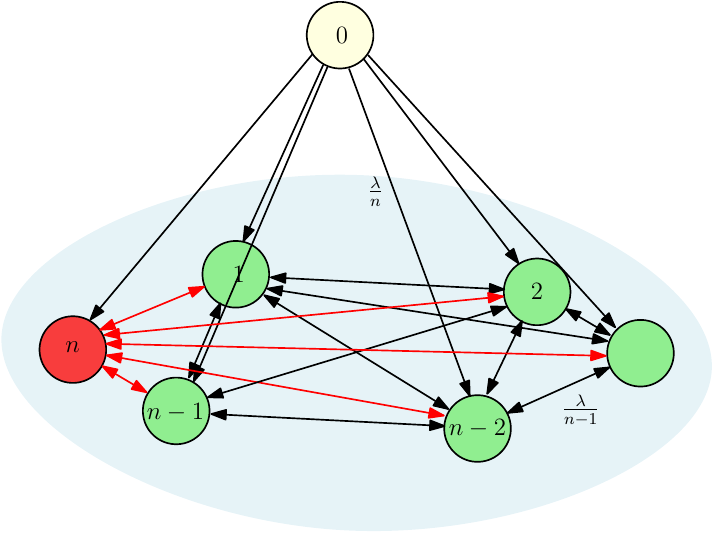}}
    \subfigure[]{\includegraphics[width=0.49\linewidth]{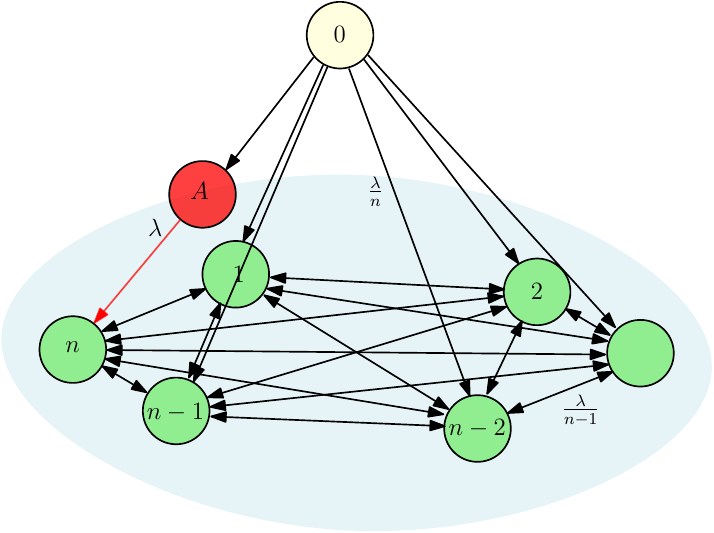}}
    \end{center}
    \caption{(a) Fully connected network of $n$ nodes with an infected node. (b) MITM attack on fully connected network of $n$ nodes.}
    \label{fig:simulation_timestomp}
\end{figure}

Some interesting results follow in case of the fully connected network shown in Fig.~\ref{fig:simulation_timestomp}(a), where it is shown that one infected node can single-handedly suppress the availability of fresh information in the entire network and increase the expected age from $O(\log n)$ found in \cite{Yates21gossip_traditional} to $O(n)$. This is achieved by increasing the timestamp of every outgoing packet to $t$ and decreasing the timestamp of every incoming packet to $0$, in effect, preventing all incoming files from being accepted by the infected node and actively persuading other nodes to always accept outgoing packets from the infected node. Additionally, if the malicious node contacts only a single node instead of all nodes of the network, see Fig.~\ref{fig:simulation_timestomp}(b), the system age still gets degraded to $O(n)$. These observations show how fully connected nature of a network can be both a benefit and a detriment for information freshness; full connectivity facilitates rapid information dissemination but, at the same time, accelerates the dissipation of adversarial inputs. On the other hand, for the unidirectional ring network, which falls on the other end of the network connectivity spectrum, it is shown that the timestomping effect on age scaling of a node is limited by its distance from the adversary, and the age scaling for a large fraction of the network continues to be $O(\sqrt{n})$, unchanged from the case with no adversary. 

The analysis involves considering an SHS, where due to the presence of a timestomping adversary, the continuous state is chosen as $(\pmb{X}(t),\pmb{U}(t))\in \mathbb{R}^{2n}$, where $\pmb{X}(t)=[X_1(t),\ldots,X_n(t)]$ denotes the instantaneous ages at the $n$ nodes and $\pmb{U}(t)=[U_1(t),\ldots,U_n(t)]$ denotes the timestamps marked on the packets at the $n$ nodes at time $t$. The state evolves in single discrete mode with differential equation $(\pmb{\dot X}(t),\pmb{\dot U}(t))=(\pmb{1}_n,\pmb{0}_n)$, as the age at each node grows at unit rate in absence of update transition and the timestamps of the node packets, both true and claimed, do not change between such transitions. Here, the actual instantaneous age at node $i$ is $X_i(t)= t-\bar{U}_i(t)$, where  $\bar{U}_i(t)$ indicates the true packet generation time, which can be different from the claimed timestamp $U_i(t)$ if the file timestamp has been tampered with. The analysis in \cite{kaswan_timestomp_jrl} primarily relies on definition of $X_{N(S)}(t)$ for a set of nodes $S$ at time $t$, which indicates the actual instantaneous age of the node claiming to possess the most recent timestamped packet in set $S$, i.e., $X_{N(S)}(t)=X_{\arg \max_{j\in S} U_j(t)}(t)$. This definition is used in \cite{kaswan_timestomp_jrl} to formulate a series of interesting test functions to explain age scaling in the different system models considered.

\section{Non-Poisson Updating}\label{sec: non-poisson}
\cite{Kaswan23aoi_nonpoisson, Kaswan23aoi_nonpoissonversion} consider timeliness in multi-hop cache-enabled networks, where updates on each link do not necessarily occur according to a Poisson process. That is, all nodes forward updates according to an ordinary renewal process, independently of other nodes. As a result, the evolution of age in the network is defined by a superposition of multiple independent ordinary renewal processes. \cite{samuels1974superpos} showed that the superposition of two ordinary renewal processes is an ordinary renewal process only if all processes are Poisson processes. Therefore, the age of information literature relies heavily on Poisson process based updates or restricts to single hop cache-updating systems, since the exponential distribution (or geometric distribution) is the only continuous (or discrete) probability distribution with memoryless property. The particular convenience of Poisson updates in SHS stems from the fact that the exponential distribution is the only distribution with constant hazard rate, which simplifies expressions involving expectations of products of test functions and transition intensities or hazard rates.
\cite{Kaswan23aoi_nonpoisson, Kaswan23aoi_nonpoissonversion} show how when the inter-update times follow general distributions which are not exponential, it becomes difficult to calibrate the different renewal processes with respect to each other, for both traditional age and version age metrics.

\begin{figure}[t]
\centerline{\includegraphics[width=0.7\linewidth]{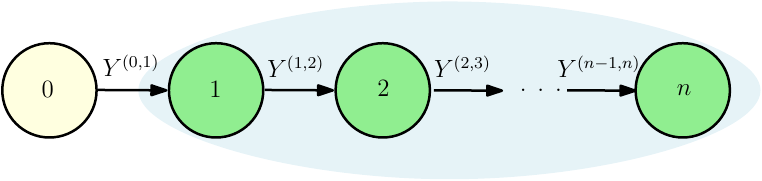}}
\caption{$n$-hop model.}
\label{fig:n_hop_model}
\end{figure}

Non-Poisson cache updating is studied through the lens of renewal theory in these works. Packets are assumed to arrive from node $i$ at node $j$ on link $(i,j)$ according to  $N^{(i,j)}(t)$ renewal process, and the finite random times $0\leq T^{(i,j)}_1 \leq T^{(i,j)}_2 \leq \ldots$ denote the corresponding renewal times, such that, inter-arrival times $Y^{(i,j)}_n=T^{(i,j)}_n- T^{(i,j)}_{n-1}$ are i.i.d.~with common distribution $F^{i,j}$. Given $N^{(i,j)}(t)=\max\{n:T^{(i,j)}_n\leq t\}$, the regenerative process $A^{(i,j)}(t)=t-T^{(i,j)}_{N^{(i,j)}(t)}$ denotes the corresponding backward recurrence time at $t$, i.e., time elapsed since the last renewal prior to $t$. \cite{Kaswan23aoi_nonpoisson} focuses on expected age in a multi-hop network of $n$ nodes as shown in Fig.~\ref{fig:n_hop_model}, and shows that the age at node $n$ is, 
\begin{align}\label{eqn: multihop age}
X_n(t)=\sum_{j=1}^{n}\Delta_j(t),
\end{align}
where 
\begin{align}\label{eqn: individual_hop age}
    \Delta_{i}(t)&=A^{(n-i,n-i+1)}\Big(t-\sum_{j=0}^{i-1}\Delta_j(t)\Big).
\end{align}
Then, (\ref{eqn: multihop age}) in conjunction with certain asymptotic results, gives
\begin{align}\label{eqn: nhop_limit_expec_age_node_n}
    \lim_{t \to \infty}\mathbb{E}[X_n(t)]=&\sum_{j=1}^{n}\lim_{t \to \infty}\mathbb{E}[\Delta_j(t)]\\
    =&\sum_{j=1}^{n}\frac{\mathbb{E}\left[\left(Y^{(n-j,n-j+1)}\right)^2\right]}{2\mathbb{E}\left[Y^{(n-j,n-j+1)}\right]}.
\end{align}

We see that the age at node $n$ depends on independent contributions of the intermediate links $(i,i+1)$, $0\leq i \leq n-1$, and is invariant to ordering of these links. Hence, each node can minimize its age by optimizing its individual packet request renewal process, irrespective of the statistical properties of other nodes and links in the network.

\cite{Kaswan23aoi_nonpoissonversion} shows an analogous additive result for version age of information in an $n$-hop network where source gets updated according to an ordinary renewal process with typical inter-renewal interval $Y^{(0,0)}$, such that version age at end user is
\begin{align}\label{eqn: nhop_limit_expec_age_node_n_V}
    \lim_{t \to \infty}\mathbb{E}[X_n(t)]=\frac{1}{\mathbb{E}[Y^{(0,0)}]}\sum_{j=1}^{n}\frac{\mathbb{E}\left[\left(Y^{(n-j,n-j+1)}\right)^2\right]}{2\mathbb{E}\left[Y^{(n-j,n-j+1)}\right]}.
\end{align}

\section{Gossiping with an Energy Harvesting Sensor}\label{sec: energy_harvesting}
\begin{figure}[t]
\centerline{\includegraphics[scale=0.8]{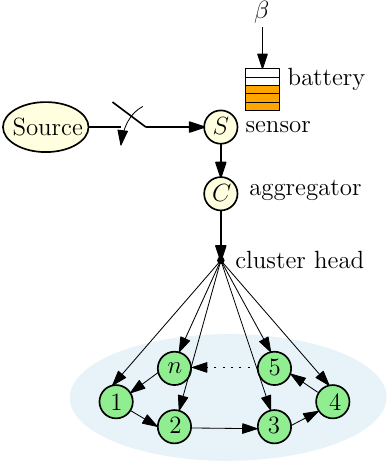}}
\caption{A gossip network with $n$ nodes and an energy harvesting sensor.}
\label{fig: MDP_system_model}
\end{figure}

\cite{Delfani2023} looks at version age in a new light in a discrete time system. It considers a sensor which can sample a source at discrete time instants and has a finite-sized battery that charges according to a Bernoulli process. There is a caching aggregator that has one storage unit and stores versions of the update. Finally, there is a gossiping uni-directional ring network of $n$ nodes; see Fig.~\ref{fig: MDP_system_model}. At any time slot, at most one node asks for updates from the aggregator. The aggregator can then send the update it has stored or ask for a fresh update from the sensor. If the sensor is asked for an update, then it samples the source and consumes one unit of battery if the battery is not empty. Node $i$ in the gossip network has $q_i$ probability of asking for an update from the aggregator. The probability that no node asks for an update in a particular time slot is $1 - \sum_{i=1}^{n}q_i$. The source updates itself with probability $p_t$ as a function of time in every time slot. The problem is to find the optimal causal policy for the aggregator to ask for updates from the source. 

The solution to this problem is found using a Markov decision process (MDP) formulation, which is different from the SHS formulation considered in most other papers with version age in gossiping networks. An MDP consists of the tuple $(S,A,P,C)$, where $S$ is the state space, $A$ is the set of
actions, $P$ is the state transition probability function, and $C$
is the cost associated with the MDP. In the context of this problem, the state at time $t$ is represented by $s(t)=\{b(t), X_1(t), X_2(t),\cdots, X_n(t), X_C(t)\}$, where $b(t)$ is the battery state, $X_i(t)$ is the version age of the $i$th node and $X_C(t)$ is the version age of the aggregator. The action at time $t$, $a(t)\in\{0,1\}$ represents the decision of the aggregator $C$ if requests for a new
update from sensor $S$, or serves the external request with a cached update. The transition probabilities $P(s'|s,a)$ describe the probability of transitioning to state $s'$ in the next timeslot, given the system is in state $s$ currently and the action taken is $a$. The update policy $\pi$, among the set of all possible policies $\Pi$, dictates the action to be taken at each timeslot. A desired policy is the one that minimizes the cost $C(s(t),a(t))=X_{avg}(t)=\frac{1}{n}\sum_{i=1}^{n}X_i(t)$, averaged over time. Hence, the optimization problem can be expressed as
\begin{align}\label{eqn: MDP_opt}
    \min_{\pi\in\Pi}\limsup_{T\to\infty}\frac{1}{T}\mathbb{E}\left[\sum_{t=0}^{T-1}X_{avg}^{\pi}(t)\bigg|s(0)=s_0\right].
\end{align}
Solving \eqref{eqn: MDP_opt}, \cite{Delfani2023} shows that the optimal policy is a threshold policy that depends only on the energy left in the battery and the version age of the aggregator; it does not depend on the version ages of the nodes in the network. 

\section{Open Problems: Future Directions}\label{sec: future_directions}
The works discussed throughout this survey article lay the foundations for tackling various aspects of the problems related to timeliness in gossip networks. However, there is an abundance of interesting open research directions. Some of these directions are discussed next. 

\paragraph{Connectivity-Freshness Synergy} The two extremes of the connectivity spectrum are, the ring network on the one end where each node is connected only to two of its neighbors, and the fully connected network on the other end where each node is connected to all of the remaining nodes. The age in the case of the ring scales as $n^{\frac{1}{2}}$ and the age in the case of the fully connected network scales as $\log n$. There is a whole range of levels of connectivity between these two extremes. At the moment, only one more connectivity point is understood in between, which is the grid network, where each node is connected to four of its neighbors. The age of a grid network scales as $n^{\frac{1}{3}}$, which is in between the age scalings of the ring and fully connected networks. This hints at a synergy between connectivity and freshness: the more connected a network is, the fresher its nodes will be through gossiping. Describing this synergy in its entirety by finding age scalings of networks with medium-level connectivities is an important open problem. 

\paragraph{Connectivity-Adversarial Action Trade-Off} While connectivity enables fast dissemination of useful information leading to better freshness for the network, it unfortunately enables fast dissemination of adversarial actions as well, such as timestomping attacks and spread of misinformation. Therefore, there may be a trade-off between connectivity of the network versus the adversarial robustness of the network. This means that there may be a trade-off between freshness of information versus spread of misinformation. Since timeliness may emerge as an important semantic feature for future high-connectivity applications, and gossiping may prove to be a simple-to-implement distributed freshness mechanism in such networks, this may open up a whole unprotected attack surface for adversarial actions, which needs to be studied carefully. To that end, it is important to study effects of connectivity on potential adversarial actions, and identify trade-offs between age of information and spread of misinformation. In addition, encryption, privacy-preserving mechanisms and physical layer security, may be integrated into age and gossiping studies. 

\paragraph{Semantics and Multi-Objective Gossiping} Gossiping so far considers only the ages of packets in spreading information in the network. We view this as a \emph{scalar} type gossiping mechanism that considers only a single variable -- the age. Future networks will consider \emph{semantics} of information, where information may have multiple features, age being one of them. Such semantics will be carried in (encoded into) different dimensions of a vector for each update packet. In such a scenario, different nodes of the network may assign priority to different dimensions, such that, for example, a node caring about the age dimension can run an age-based gossiping protocol, while another node caring about another dimension of the packet may resort to a different gossiping protocol. Studying multi-objective gossiping, and considering synergies and trade-offs between different semantic components of a gossiping vector, are future research directions.
    
\paragraph{Mobility and Time-Varying Connections in Gossiping} The current literature only considers gossiping networks which are static. Mobility and time-varying connections will be part of future dense networks such as vehicular networks, robotic networks, and drone swarms. Mobility will constantly change the connectivity and topology of such networks. It is well-known that mobility increases the capacity (throughput) of wireless networks. It would be interesting to study the effects of mobility on the freshness of information in networks. 
    
\paragraph{Tracking Multiple Sources via Gossiping} So far, we have considered networks that monitor and update the versions of a single observable (i.e., samples of a single random process). In future networks, multiple observables or samples of multiple random processes will flow through a network. In such a system, where each node possesses various samples of multiple information sources, when two nodes get in touch and have an opportunity to gossip, which information should they pass on: Should they randomly choose one of the sources and gossip the freshest sample they have from that source? Should they go over the sources in a round robin fashion? Instead of sending the freshest sample of a single information source, should they linearly combine samples from different information sources and gossip that \emph{mixed} sample? These are interesting open problems that can be studied. In addition, different approaches may be needed depending on whether these multiple information sources are independent or correlated. 

\section{Conclusion}\label{sec: conclusions}
This article provides a comprehensive overview of gossiping as a communication mechanism for rapid information dissemination in networks, with emphasis on use of age of information as a key metric. The overview covers the evolution of gossip algorithms, starting from dissemination of static messages, all the way to real-time gossiping. The evaluation of gossip timeliness scaling across different network topologies, including fully connected, ring, grid, generalized ring, hierarchical, and sparse asymmetric networks, is discussed. The article explores age-aware gossiping, higher-order moments of the age process, and various network variations such as file slicing, network coding, reliable and unreliable sources, information mutation, adversarial actions, and energy harvesting sensors. In addition to consolidation of the recent developments in timely gossiping, the article also outlines several open problems and future directions, which may be of interest.

\bibliographystyle{unsrt}
\bibliography{references}

\end{document}